\documentclass[12pt]{iopart}

\usepackage{iopams}  
\usepackage{graphicx}

\newtheorem{theorem}{Theorem}

\begin{document}

\title[Quantum Fluctuations in Mesoscopic Systems]{Quantum Fluctuations in Mesoscopic Systems}

\author{F. Benatti$^{1,2}$, F. Carollo$^{3}$, R. Floreanini$^2$\, H. Narnhofer$^4$}

\address{${}^1$Dipartimento di Fisica, Universit\`a di Trieste, Trieste, 34151 Italy}
\address{${}^2$Istituto Nazionale di Fisica Nucleare, Sezione di Trieste, 34151 Trieste, Italy}
\address{${}^3$School of Physics and Astronomy and Centre for the Mathematics and Theoretical Physics of Quantum Non-Equilibrium Systems, University of Nottingham, Nottingham, NG7 2RD, UK}
\address{${}^4$Institut f\"ur Theoretische Physik, Universit\"at Wien, A-1091, Vienna, Austria}

\ead{benatti@ts.infn.it, Federico.Carollo@nottingham.ac.uk, floreanini@ts.infn.it,
heide.narnhofer@univie.ac.at}
%\vspace{10pt}
%\begin{indented}
%\item[]June 2016
%\end{indented}

\begin{abstract}
Recent experimental results point to the existence of
coherent quantum phenomena in systems made of a large number of particles, 
despite the fact that for many-body systems the presence of decoherence 
is hardly negligible and emerging classicality is expected. 
This behaviour hinges on collective observables,
named {\sl quantum fluctuations}, that retain 
a quantum character even in the thermodynamic limit:                                                       
they provide useful tools for studying properties of many-body systems at the mesoscopic level,
in between the quantum microscopic scale and the classical macroscopic one.  
We hereby present the general theory of quantum fluctuations in mesoscopic systems 
and study their dynamics in a quantum open system
setting, taking into account the unavoidable effects of dissipation and noise 
induced by the external environment. 
As in the case of microscopic systems, decoherence is not always the only dominating effect 
at the mesoscopic scale: certain type of environments can provide means 
for entangling collective fluctuations through a purely noisy mechanism.
\end{abstract}

\vspace{2pc}
\noindent{\it Keywords}: quantum fluctuations, mesoscopic systems, open quantum dynamics, entanglement
%\maketitle

\vspace{10pt}

\section{Introduction}
\label{section1}

When dealing with quantum systems formed by a large number of elementary constituents, the study of
their microscopic properties becomes impractical, due to the high multiplicity of the basic elements.
Instead, collective observables, {\it i.e.} observables involving all system degrees of freedom, can be
directly connected to measurable quantities, and therefore constitute the most suited operators
to be used to describe the physical properties of such many-body systems. 
Collective observables
are of extensive character, growing indefinitely as the number $N$ of microscopic constituents 
becomes large: they need to be normalized by suitable powers of $1/N$ 
in order to obtain physically sensible definitions.
In this way, provided the system density is kept fixed, these normalized collective
observables become independent from the number of constituents, allowing oneself to work in the
so-called thermodynamic, large $N$ limit \cite{Feynman}-\cite{Sewell1}.

Typical examples of collective observables are provided by the so-called {\it mean-field} operators:
they are averages over all constituents of single particle quantities,
an example of which is the mean magnetization in spin systems.
Although the single particle observables possess a quantum character,
mean-field observables show in general a classical-like behaviour
as the number $N$ of constituents increases,
thus becoming examples of the so-called {\it macroscopic observables}.
The well-established mean-field approach to the study of many-body systems 
precisely accounts for their behaviour at this macroscopic, semiclassical level,
where very little, if none, quantum character survives.

It thus came as a surprise the report of having observed coherent
quantum behaviour also in systems made of a large number of particles
\cite{Julsgaard}-\cite{Purdy}, typically,
involving Bose-Einstein condensates, namely thousands of ultracold atoms 
trapped in optical lattices \cite{Leggett1}-\cite{Lewenstein2}, 
hybrid atom-photon \cite{Haroche}-\cite{Klimov} 
or optomechanical systems \cite{Wallquist}-\cite{Bowen}, 
where decoherence effects can hardly be neglected and emerging
classicality is ultimately expected. 
Mean-field observables can not be used to explain such a behaviour: 
as mentioned, being averages quantities, scaling as $1/N$ for large $N$,
they show a semiclassical character. 
However, other kinds of collective observables
have been introduced and studied in many-body systems \cite{Goderis1}-\cite{Verbeure-book}; 
they account for the variation of microscopic quantities around their averages 
computed with respect to a chosen reference state:
in analogy with classical probability theory, they are called {\it quantum fluctuations}.
These observables still involve all system degrees of freedom; however, scaling as $1/\sqrt{N}$
with the number of constituents, they retain some quantum properties
even in the thermodynamic limit. Being half-way between the microscopic observables,
those describing the behaviour of single particles in the system, and the macroscopic mean-field
observables, they are named {\it mesoscopic}. Indeed,
quantum fluctuations always form noncommutative algebras, thus providing 
a useful tool for analyzing those quantum many-body properties 
that persist at an intermediate scale, 
in between the microscopic world and the classical macroscopic one. 

One of the most striking manifestation of a quantum behaviour is 
the possibility of establishing correlations between parts of a physical system
that have no classical analog, 
{\it i.e.} of generating {\it entanglement} between them \cite{Horodecki}-\cite{Modi}.
At first considered as a mere curiosity, quantum correlations and entanglement have nowadays
become physical resources allowing the realization of protocols and 
tasks in quantum information technologies
not permitted by purely classical means \cite{Nielsen,Petritis}.

Entanglement is however an extremely fragile resource, that can be rapidly depleted
by the action of an external environment. In general, any quantum system, and in particular
a many-body one, can hardly be considered to be completely isolated: coupling to its surroundings
is unavoidable and, generically, this leads to noisy and decoherence effects,
eventually washing away any quantum behaviour \cite{Alicki1}-\cite{Spohn1}.

Nevertheless, it has been found that an external environment can be responsible not only 
for degrading quantum coherence and entanglement,
but, quite surprisingly, also of enhancing quantum correlations through a purely mixing mechanism.
Indeed, it has been shown that, in certain circumstances, two independent, 
non interacting systems can become entangled by the action of a common bath 
in which they are immersed. In general, the obvious way of entangling
two quantum systems is through a direct interaction among them; a different
possibility is to put them in contact with an external environment:
the presence of the bath induces a mixing-enhancing mechanism able to
actually generate quantum correlations among them \cite{Plenio1}-\cite{Benatti5}.
This interesting effect has been proven to occur
in microscopic systems, made of two qubits or oscillators;
surprisingly, it works at the mesoscopic scale
also in many-body systems, provided one focuses on suitably chosen 
fluctuation observables.

Aim of this report is to give an overview of the theory of
quantum fluctuations in reference to
quantum correlations and entanglement in open many-body quantum systems 
at the mesoscopic scale.

Observables having the form of fluctuations were first introduced in the late 1980's
in the analysis of quantum lattice systems with short-range interactions \cite{Goderis1}-\cite{Goderis3}.
There, it was observed that the set of all these fluctuation observables form an algebra, 
that, irrespective of the nature of the microscopic constituents, 
turns out to be nonclassical, {\it i.e.} noncommutative,
and always of bosonic character: it is at the elements of this algebra that one should
look in order to properly describe quantum features of many-body systems
at the mesoscopic scale. These results 
proved to be very useful in understanding the basis of the linear response theory 
and the Onsager relations \cite{Goderis4}-\cite{Goderis6}, and started extensive studies
on the characteristics and basic time evolution 
of the fluctuation operator algebra in various physical models \cite{Goderis7}-\cite{Pascazio}.

Despite these successes, since recently very little was known
of the behaviour of quantum fluctuations in open many-body systems, {\it i.e.} in systems 
in contact with an external environment:
this is the most common situation encountered in actual experiments, that can
never be thought of as completely isolated from their surroundings.
Taking as reference systems models made of a collection of either spins or oscillators
immersed in a common bath, a comprehensive analysis
of open, dissipative dynamics of many-body fluctuation operators can be given \cite{Carollo1}-\cite{Surace}.
With respect to the unitary time evolutions explored so far,
the presence of the external environment poses specific challenges
in the derivation of the mesoscopic dynamics, leading however to interesting new
physical results: two non-interacting many-body systems in
a common bath can become entangled at the level of mesoscopic fluctuations, and,
in certain situations, the created quantum correlations can persisit 
even for asymptotic long times.

Of particular interest is the application of the theory of quantum fluctuations to models
with long-range interactions \cite{Binney}-\cite{Bagarello},
shedding new light on the physical properties of such many-body systems at the mesoscopic scale.
In these cases, the microscopic dynamics is implemented through mean-field operators, {\it i.e.}
with interaction and dissipative terms scaling as $1/N$; in the thermodynamic limit,
it converges to a non-Markovian \cite{Rivas2}-\cite{Vega}, unitary dynamics on local operators, while giving rise
to a non-linear, dissipative dynamics
at the level of quantum fluctuations.

\bigskip

In detail the structure of the review is as follows.

In the following Section, the basic mathematical tools for the description of many-body quantum systems 
are briefly reviewed: they are based on the algebraic approach to quantum mechanics, 
which represents the most general formulation of the theory, valid for both
finite and infinite dimensional systems \cite{Bratteli}-\cite{Spohn2}.
The characteristic properties of collective many-body observables,
and in particular quantum fluctuations, are subsequently discussed:
in presence of short-range correlations, in the thermodynamic limit, 
fluctuation operators are seen to become
bosonic quantum variables with Gaussian characteristic function \cite{Adesso-thesis}-\cite{Eberly}.
Such a limiting behaviour is rooted in the extension to the quantum setting
of the classical central limit theorem \cite{Cushen,Quagebeur}.
These abstract results are then applied to the discussion of many-body systems 
composed by spin-chains or collections of independent oscillators.

Section \ref{section3} is instead devoted to the study of the dynamics of quantum fluctuations.
The focus is on open, dissipative
time evolutions as given by microscopic, local generators in 
Kossakowski-Lindblad form \cite{Kossakowski1}-\cite{Lindblad}.
Under rather general conditions, one can show that the emergent, large $N$ mesoscopic dynamics
for the bosonic fluctuations turns out to be a quantum dynamical semigroup of quasi-free type,
thus preserving the Gaussian character of the fluctuation algebra.
When dealing with bipartite many-body systems, this emergent dissipative Gaussian dynamics
is able to create mesoscopic entanglement at the level of fluctuation operators
through a purely noisy mechanism, namely, without environment mediated interaction among the
mesoscopic degrees of freedom.
Remarkably, in certain situations, the generated entanglement
can persist for asymptotic long times.
The behaviour of the created collective quantum correlations can be studied
as a function of the characteristics of the external environment in which
the mesoscopic system is immersed.
One then discovers that a sort of entanglement phase transition is at work:
a critical temperature can always be identified, above which quantum correlations 
between mesoscopic observables can not be created. 

Section \ref{section4} deals with systems with long-range interactions
\cite{Thirring3}-\cite{Bagarello}.
In the thermodynamic limit, the dissipative dynamics of such systems 
behaves quite differently depending on whether one focuses on microscopic or collective observables.
Quite surprisingly, the time evolution of local, {\it i.e.} microscopic, observables turns out to be an automorphism
of non-Markovian character,
generated by a time-dependent Hamiltonian, while that of quantum fluctuations,
{\it i.e.} of mesoscopic observables,
consists of a one-parameter family of non-linear maps.
These maps can be extended to a larger algebra in such a way that their generator 
becomes time-independent, giving rise to a semigroup of completely positive maps,
whose generator is however of hybrid type, containing
quantum as well as classical contributions.

Finally, let us point out that the theory of quantum fluctuations
is very general and independent from the specific models here discussed. 
In this respect, it can be applied in all instances where mesoscopic, 
coherent quantum behaviours are expected to emerge, {\it e.g.} 
in experiments involving spin-like and optomechanical systems, 
or trapped ultra-cold atom gases: 
the possibility of entangling these many-body systems through 
a purely mixing mechanism may reinforce their use 
in the actual realization of quantum information and communication protocols.

\vskip 1cm

\section{Many-body collective observables}
\label{section2}

We shall consider quantum systems composed by $N$
(distinguishable) particles
and analyze their behavior in the the so-called thermodynamic,
large $N$ limit by studying their collective properties.

The proper treatment of infinite quantum systems requires
the use of the algebraic approach to quantum physics: 
in the coming subsection, we shall briefly summarize 
its main features, underlying the concepts and tools that will be needed
in the following discussions. [For a more detailed presentation, see the reference textbooks 
\cite{Bratteli}-\cite{Strocchi2}.]

\subsection{Observables and states}
\label{section2.1}

Any quantum system can be characterized by the collections of observations
that can be made on it through suitable measurement processes \cite{Strocchi3}. The physical quantities
that are thus accessed are the observables of the system, forming an algebra $\cal A$
under multiplication and linear combinations, the algebra of observables.

\bigskip
\noindent
$\bullet$ {\sl $C^\star$-algebras}\hfill\break
In general, the algebra $\cal A$ turns out to be a non-commutative $C^\star$-algebra; 
this means that it is a linear,
associative algebra (with unity) over the field of complex numbers $\mathbb{C}$,
{\it i.e.} a vector space over $\mathbb{C}$, with an associative product, linear
in both factors. Further, $\cal A$ is endowed with an operation of conjugation:
it posses an antilinear involution $\star: {\cal A}\to {\cal A}$, such that
$(\alpha^\star)^\star=\alpha$, for any element $\alpha$ of $\cal A$.
In addition, a norm $|| \cdot ||$ is defined on $\cal A$, satisfying
$||\alpha\beta||\leq ||\alpha||\, ||\beta||$, for any $\alpha,\, \beta \in {\cal A}$
(thus implying that the product operation is continuous), and such that
$||\alpha^\star \alpha||=||\alpha||^2$, so that $||\alpha^\star||=||\alpha||$;
moreover, $\cal A$ is closed under this norm, meaning that $\cal A$ is a complete
space with respect to the topology induced by the norm (a property that in turn makes $\cal A$
a Banach algebra).

In the case of an $n$-level system, $\cal A$ can be identified with the $C^\star$-algebra
${\cal M}_n(\mathbb{C})$ of complex $n\times n$ matrices; the $\star$-operation
coincides now with the hermitian conjugation, $M^\star=M^\dagger$, 
for any element $M\in {\cal M}_n(\mathbb{C})$,
while the norm $||M||$ is given by the square root of the largest eigenvalue of $M^\dagger M$.
Nevertheless, the description of a physical system through its $C^\star$-algebra of observables
is particularly appropriate in presence of an infinite number of degrees of freedom,
where the canonical formalism is in general problematic.

\bigskip
\noindent
$\bullet$ {\sl States on $C^\star$-algebras}\hfill\break
Although the system observables, {\it i.e.} the hermitian elements of $\cal A$, can be
identified with the physical quantities measured in experiments, the explicit link between
the algebra $\cal A$ and the outcome of the measurements is given by the concept of a state
$\omega$, through which the expectation value $\omega(\alpha)$ of the observable $\alpha \in {\cal A}$
can be defined.

In general, a state $\omega$ on a $C^\star$-algebra $\cal A$ is a linear map
$\omega: {\cal A} \to \mathbb{C}$, with the property of being positive,
{\it i.e.} $\omega(\alpha^\star \alpha)\geq 0$, $\forall\alpha \in {\cal A}$,
and normalized, $\omega({\bf 1})=1$, indicating with ${\bf 1}$ the unit of $\cal A$.
It immediately follows that the map $\omega$ is also continuous:
$|\omega(\alpha)|\leq ||\alpha||$, for all $\alpha \in {\cal A}$.

This general definition of state of a quantum system comprises the standard one
in terms of normalized density matrices on a Hilbert space $\cal H$;
indeed, any density matrix $\rho$ defines a state $\omega_\rho$ on the algebra ${\cal B}({\cal H})$
of bounded operators on ${\cal H}$
through the relation
\begin{equation}
\omega_\rho(\alpha)={\rm Tr}[\rho\,\alpha]\ ,\qquad \forall\alpha\in {\cal B}({\cal H})\ ,
\label{2.1}
\end{equation}
which for pure states, $\rho=|\psi\rangle\langle\psi|$, reduces to the
standard expectation: $\omega_\rho(\alpha)=\langle \psi |\alpha|\psi\rangle$.
Nevertheless, the definition in terms of $\omega$ is more general, holding
even for systems with infinitely many degrees of freedom, for which
the usual approach in terms of state vectors may be unavailable.

As for density matrices on a Hilbert space $\cal H$, a state $\omega$ on a $C^\star$-algebra $\cal A$
is said to be pure if it can not be decomposed as a convex sum of two
states, {\it i.e.} if the decomposition $\omega=\lambda\,\omega_1+(1-\lambda)\, \omega_2$,
with $0\leq\lambda\leq 1$, holds only for $\omega_1=\omega_2=\omega$. If a state $\omega$
is not pure, it is called mixed. It is worth noticing that, for consistency, the assumed
completeness of the relation between observables and measurements on a physical system
requires that the observables separate the states, {\it i.e.} $\omega_1(\alpha)=\omega_2(\alpha)$
for all $\alpha \in {\cal A}$ implies $\omega_1=\omega_2$, and similarly that the states separate
the observables, {\it i.e.} $\omega(\alpha)=\omega(\beta)$ for all states $\omega$ on $\cal A$
implies $\alpha=\beta$.

\bigskip
\noindent
$\bullet$ {\sl GNS-Construction}\hfill\break
Although the above description of a quantum system through its $C^\star$-algebra of observables
(its measurable properties) and states over it (giving the observable expectations)
looks rather abstract, it actually allows an Hilbert space interpretation, through the so-called
{\it Gelfang-Naimark-Segal(GNS)-construction}.

\medskip
\noindent
{\bf GNS Theorem} {\it Any state $\omega$ on the $C^\star$-algebra $\cal A$
uniquely determines (up to isometries) a representation $\pi_\omega$ of the elements of $\cal A$
as operators in a Hilbert space ${\cal H}_\omega$, containing a reference vector
$|\omega\rangle$, whose matrix elements reproduce the observable expectations:
\begin{equation}
\omega(\alpha)=\langle\omega| \pi_\omega(\alpha) |\omega\rangle\ ,\qquad \alpha\in{\cal A}\ .
\label{2.2}
\end{equation}
}

\noindent
This result makes apparent that the notion of Hilbert space associated to a quantum system
is not a primary concept, but an emergent tool, a consequence of the $C^\star$-algebra structure
of the system observables. We shall now apply these basic algebraic tools to the description
of many-body quantum systems.

\subsection{Quasi-local algebra}
\label{section2.2}

Being distinguishable, each particle in the many-body system can be identified 
by an integer index $k\in\mathbb{N}$. In view of the previous discussion,
its physical properties can be described by the $C^\star$ algebra $\mathfrak{a}^{[k]}$ of single-particle
observables, that will be assumed to be the same algebra $\mathfrak{a}$ for all particles. When its
dimension $d$ is finite, $\mathfrak{a}$ can be identified with $\mathcal{M}_d(\mathbb{C})$;
nevertheless, it can be also infinite-dimensional ({\it e.g.} the oscillator algebra).

Referring to different degrees of freedom, operator algebras of different particles commute: 
$\left[\mathfrak{a}^{[i]},\mathfrak{a}^{[j]}\right]=0$, $i\neq j$.
By means of the tensor product structure one can construct {\it local algebras}, 
referring just to a finite number of particles. For instance, the algebra
\begin{equation}
\mathcal{A}_{[p,q]}=\bigotimes_{i=p}^{q}\mathfrak{a}^{[i]},\qquad p,q\in\mathbb{N},\ p\le q \ ,
\label{2.3}
\end{equation}
contains all observables pertaining to the set of particles whose label is between $p$ and $q$. 
The family of local algebras $\left\{\mathcal{A}_{[p,q]}\right\}_{p\le q}$ possesses the following properties \cite{Bratteli}:
\begin{eqnarray}
\nonumber
\Big[\mathcal{A}_{[p_1,q_1]},\, \mathcal{A}_{[p_2,q_2]}\Big]=0\qquad {\rm if}\ \ [p_1,q_1]\cap [p_2,q_2]=\emptyset\ ,\\
\mathcal{A}_{[p_1,q_1]}\subseteq\mathcal{A}_{[p_2,q_2]}\qquad\qquad {\rm if }\ \ [p_1,q_1]\ \subseteq [p_2,q_2]\ .
\nonumber
\end{eqnarray}
One then consider the union of these algebras over all possible finite sets of particles, 
$\bigcup_{p\le q}\mathcal{A}_{[p,q]}$, and its completion with respect to the norm inherited 
from the local algebras. The resulting algebra $\mathcal{A}$ is called the {\it quasi-local} algebra: it
contains all the observables of the system.  
In the following, generic elements of $\mathcal{A}$ will be denoted with capital letters, $X$,
while lower case letters, $x$, will represent elements of $\mathfrak{a}$. Actually,
any observable $x\in \mathfrak{a}$ of particle $k$ can be embedded into $\mathcal{A}$ as
\begin{equation}
x^{[k]}=\ldots \otimes{\bf 1}\otimes x\otimes {\bf 1}\otimes\ldots\ ,
\label{2.4}
\end{equation}
where in the above infinite tensor product of identity operators, $x$ appears exactly at position $k$.
As a result, $x^{[k]}\in\mathcal{A}$ acts non trivially only on the $k$-th particle.
Furthermore, some operators in the quasi-local algebra $\mathcal{A}$ act non-trivially 
only on a finite set of particles: they will be called (strictly) {\it local} operators.
Since $\mathcal{A}$ is the norm closure of the union of all possible local algebras, the set
of all its local elements is dense; in other terms, any element of $\mathcal{A}$
can be approximated (in norm) by local operators, with an error that can be made arbitrarily small.

States for the system will be described by positive, normalized, linear functionals $\omega$ on $\mathcal{A}$:
they assign the expectation value $\omega(X)$ to any operator $X\in\mathcal{A}$. In the following, we shall restrict
the attention to states for which the expectation values of a same observable for different particles coincide:
\begin{equation}
\omega\big(x^{[j]}\big)=\omega\big(x^{[k]}\big)\ ,\qquad j\neq k\ .
\label{2.5}
\end{equation}
In other terms, the mean value of single-particle operators are the same for all particles;
to remark this fact, we shall use the simpler notation:
\begin{equation}
\omega\big(x^{[k]}\big)\equiv\omega(x)\ ,\qquad x\in\mathfrak{a}\ .
\label{2.6}
\end{equation}
When the single-particle algebra $\mathfrak{a}$ is finite dimensional, recalling (\ref{2.1}), one can
further write: $\omega(x)=\Tr[\rho\, x]$, with $\rho$ a single-particle density matrix.

In addition to property (\ref{2.5}), called {\it translation invariance}, 
we shall require that the states $\omega$ of the system to be also {\it clustering}, {\it i.e.}
not supporting correlations between far away localized operators:
\begin{equation}
\hskip -.5cm
\lim_{|z|\to\infty}\omega\Big(A^\dagger\, \tau_z\big(X\big)\, B\Big)=
\omega\Big(A^\dagger\, B\Big)\, \lim_{|z|\to\infty}\omega\Big(\tau_z\big(X\big)\Big)
=\omega\Big(A^\dagger\, B\Big)\,\omega\big(X\big)\ ,
\label{2.7}
\end{equation}
where $\tau_z:\mathcal{A}\to\mathcal{A}$ is the spacial translation operator. 

Using this algebraic setting,
we shall see that the common wisdom that assigns a ``classical'' behaviour
to operator averages while a non-trivial dynamics to fluctuations holds also in the case
of quantum many-body systems. More specifically, mean-field observables will be shown to provide 
a classical (commutative) description of the system, typical of the ``macroscopic'' world, 
while fluctuations around operator averages will still retain some quantum (noncommutative) properties: 
they describe the ``mesoscopic'' behaviour of the system, at a level that is half way between the
microscopic and macroscopic scale.

\subsection{Mean-field observables}
\label{section2.3}

Single-particle operators, or more in general local operators, are observables suitable for a microscopic
description of a many-body system. However, due to experimental limitations, these operators are hardly 
accessible in practice; only, collective observables, involving all system particles, are in general
available to the experimental investigation. 

In order to move from a microscopic description to a one involving collective operators, potentially
defined over system with an infinitely large number of constituents, a suitable scaling needs to be chosen.
The simplest example of collective observables are {\it mean-field} operators, 
{\it i.e.} averages of $N$ copies of a same single site observable $x$:
\begin{equation}
\overline{X}^{(N)}=\frac{1}{N}\sum_{k=1}^N x^{[k]}\ .
\label{2.8}
\end{equation}
We are interested in
studying their behaviour in the thermodynamic, large $N$ limit.

As a first, preliminary step, let us consider two such operators, $\overline{X}^{(N)}$ and $\overline{Y}^{(N)}$, 
constructed from single-particle observables $x$ and $y$,
respectively, and compute their commutator:
\begin{equation}
\Big[\overline{X}^{(N)},\, \overline{Y}^{(N)}\Big]= \frac{1}{N^2}\sum_{j,k=1}^N \Big[x^{[j]},\, y^{[k]}\Big]=
\frac{1}{N^2}\sum_{k=1}^N \Big[x^{[k]},\, y^{[k]}\Big]\ ,
\label{2.9}
\end{equation}
where the last equality comes from the fact that operators referring to different particles commute.
Since $(1/N)\sum_{k=1}^N \Big[x^{[k]},\, y^{[k]}\Big]$ is clearly itself a a mean-field operator,
one realizes that the commutator of two mean-field operators is still a mean-field operator,
although with an additional $1/N$ factor; because of this extra factor, it vanishes in the large $N$ limit.
In other terms, mean-field operators seem to provide only a ``classical'', 
commutative description of the many-body system,
any quantum, non-commutative character being lost in the thermodynamic limit.

The above result actually holds in the so-called {\it weak operator topology} \cite{Bratteli}, 
{\it i.e.} under state average. More precisely, for a clustering state $\omega$, one has:
\begin{equation}
\lim_{N\to\infty}\omega\left(A^\dagger\, \overline{X}^{(N)}\, B\right)=\omega\big(A^\dagger B\big)\,\omega(x)\ ,
\qquad A,\, B\in\mathcal{A}\ .
\label{2.10}
\end{equation}
Indeed, for any integer $N_0<N$ one can write:
$$
\lim_{N\to\infty} \omega\left(A^\dagger\, \overline{X}^{(N)}\, B\right)=
\lim_{N\to\infty} \omega\Bigg( A^\dagger\ \Bigg[ \frac{1}{N} \sum_{k=1}^{N_0} x^{[k]} 
+ \frac{1}{N} \sum_{k=N_0+1}^{N} x^{[k]}\Bigg]\, B\Bigg)\ .
$$
Clearly, the first piece in the r.h.s. gives no contributions in the limit. 
Concerning the second term, we can appeal to the fact that
local operators are norm dense in $\mathcal{A}$; then, without
loss of generality, one can assume $N_0$ to be large so that $B$ involves only particles with labels $\leq N_0$.
Recalling the clustering property (\ref{2.7}),
one then immediately gets the result (\ref{2.10}). This means that, in the weak operator topology,
the large $N$ limit of $\overline{X}^{(N)}$ is a scalar multiple of the identity operator:
$$
\lim_{N\to\infty} \overline{X}^{(N)} = \omega(x)\, {\bf 1}\ .
$$
With similar manipulations, one can also prove that the product $\overline{X}^{(N)}\overline{Y}^{(N)}$ 
of two mean-field-observables
weakly converges to $\omega(x)\omega(y)$ \cite{Carollo3}: 
\begin{equation}
\lim_{N\to\infty}\overline{X}^{(N)}\,\overline{Y}^{(N)}\,=\,\omega(x)\,\omega(y)\, {\bf 1}\ .
\label{2.11}
\end{equation}
Furthermore, under the stronger $L_1$-clustering condition (see next Section and~\cite{Verbeure-book}),
\begin{equation}
\sum_{k\in\mathbb{N}}\left|\omega\big(x^{[1]}y^{[k]}\big)\,-\,\omega(x)\omega(y)\right|<\infty\, ,
\label{2.12}
\end{equation}
the following scaling can be proven \cite{Carollo3}:
\begin{equation}
\label{2.13}
\left|\omega\Big( \overline{X}^{(N)}\,\overline{Y}^{(N)}\Big)-\,\omega(x)\,\omega(y)\right|
=O\left(\frac{1}{N}\right)\, .
\end{equation}
It thus follows that the weak-limit of mean-field observables gives rise to
a commutative (von Neumann) algebra. 

Therefore, mean-field observables describe what we can call ``macroscopic'', classical degrees of freedom;
although constructed in terms of microscopic operators, in the large $N$ limit they do not retain
any fingerprint of a quantum behaviour. Instead, as remarked in the Introduction, we are interested in
studying collective observables, {\it} involving all system particles, showing a quantum
character even in the thermodynamic limit. Clearly, a less rapid scaling than $1/N$ is needed.

\subsection{Quantum fluctuations}
\label{section2.4}

Fluctuation operators are collective observables that scale as the square root of $N$ and represent a deviation
from the average. Given any single-particle operator $x$ and a reference state $\omega$, its corresponding
fluctuation operator $F^{(N)}(x)$ is defined as
\begin{equation}
F^{(N)}(x)\equiv \frac{1}{\sqrt N}\sum_{k=1}^N \Big(x^{[k]}-\omega(x){\bf 1}\Big)\ ;
\label{2.14}
\end{equation}
it is the quantum analog of a fluctuation random variable in classical probability theory \cite{Feller}.

Although the scaling $1/\sqrt N$ does not in general guarantee convergence in the weak operator topology, 
one can make sense of the large $N$ limit of (\ref{2.14}) in some state-induced topology.
Indeed, note that the mean value of the fluctuation always vanishes: $\omega\big(F^{(N)}(x)\big)=\,0$.
Moreover, one has:
\begin{eqnarray}
\nonumber
\lim_{N\to\infty}\omega\Big(\big[F^{(N)}(x)\big]^2\Big)=
\lim_{N\to\infty}\frac{1}{N}\sum_{j,k=1}^N\Big(\omega\left(x^{[j]}x^{[k]}\Big)-\omega(x)^2\right)\\
\hskip 7cm \le\sum_{k\in\mathbb{N}}\Big|\omega\left(x^{[1]}x^{[k]}\right)-\omega(x)^2\Big|\ ,
\label{2.15}
\end{eqnarray}
so that for states satisfying the $L_1$-clustering condition, introduced earlier in (\ref{2.12}),
the variance of the fluctuations is bounded in the limit of large $N$.

In addition,
fluctuation operators retain a quantum behaviour in the large $N$ limit. 
Consider two single-particle operators $x,\, y\in \mathfrak{a}$ and call $z\in \mathfrak{a}$ their commutator.
Since $[x^{[j]}\,,\, y^{[k]}]=\delta_{jk}\,z^{[j]}$, following steps similar to the one used
in the proof of (\ref{2.10}), one can write for a clustering state $\omega$: 
$$
\lim_{N\to\infty} \omega\left(A^\dagger\ \left[F^{(N)}(x)\,,\,F^{(N)}(y)\right]\, B\right)=\lim_{N\to\infty} 
\frac{1}{N}\sum_{k=1}^{N}\omega\left(A^\dagger z^{(k)}\, B\right)=\omega(A^\dagger B)\,\omega(z) ,
$$
with $A$ and $B$ arbitrary elements of $\mathcal{A}$.
Thus, commutators of fluctuations of local operators give rise to mean-field observables,
and as such, behave for large $N$ as scalar multiples of the identity, $\omega(z)\,\mathbf{1}$.
In other terms, in the thermodynamic limit fluctuations provide commutation relations
that look like standard canonical bosonic ones.
These results indicate that, at the mesoscopic level, a  non-commutative
bosonic algebraic structure naturally emerges: 
quantum fluctuations indeed form a so-called {\it quantum fluctuation algebra}.

In order to explicitly construct this algebra, one starts by considering 
the set of self-adjoint elements of the quasi-local algebra $\mathcal{A}$. Actually, as shown
by the examples presented below, only subsets of this set are in general 
physically relevant, so that one can limit the discussion to one of them.
Let us then fix a set of linearly independent, self-adjoint elements
$\{x_1,\ x_2,\ \ldots ,\ x_n\}$ in the single-particle algebra $\mathfrak{a}$ and consider their real linear span:
\begin{equation}
\mathcal{X}=\Big\{x_r\ \big|\ x_r\equiv\vec{r}\cdot\vec{x}=\sum_{\mu=1}^n r_\mu\, x_\mu,\ \vec{r}\in\mathbb{R}^n\Big\}\ .
\label{2.16}
\end{equation}
Following the definition (\ref{2.14}), one can then construct the fluctuation operators $F^{(N)}(x_\mu)$ corresponding to $x_\mu$, $\mu=1,2,\ldots, n$, and the one corresponding to the generic combination $x_r\in\mathcal{X}$,
obtained from those by linearity:
\begin{equation}
F^{(N)}(x_r)=\sum_{\mu=1}^N r_\mu\, F^{(N)}(x_\mu)\equiv \vec{r}\cdot\vec{F}^{(N)}(x)\ ;
\label{2.17}
\end{equation}
we want to study the large $N$ behaviour of these fluctuation operators, having fixed a state $\omega$
satisfying the invariance and clustering properties in (\ref{2.5}) and (\ref{2.7}).

In order to build well behaved fluctuations, the discussion leading to (\ref{2.15}) suggests to choose
observables $x_\mu$ for which the $L_1$-clustering property (\ref{2.12}) is satisfied 
for all elements of the space $\mathcal{X}$. This condition guaranties that the $n\times n$ {\it correlation matrix} 
$C^{(\omega)}$, with components:
\begin{equation}
C^{(\omega)}_{\mu\nu}=\lim_{N\to\infty} \omega\Big( F^{(N)}(x_\mu)\ F^{(N)}(x_\nu)\Big)\ ,
\qquad \mu,\nu=1,2,\ldots, n\ ,
\label{2.18}
\end{equation}
be well defined \cite{Verbeure-book}. This matrix can be decomposed as
\begin{equation}
C^{(\omega)}=\Sigma^{(\omega)} + \frac{i}{2} \sigma^{(\omega)}\ ,
\label{2.19}
\end{equation}
in terms of the {\it covariance matrix}, namely its real, symmetric part $\Sigma^{(\omega)}$, with components
\begin{equation}
\Sigma^{(\omega)}_{\mu\nu}=\frac{1}{2}\lim_{N\to\infty} \omega\Big( \Big\{F^{(N)}(x_\mu),\ F^{(N)}(x_\nu)\Big\}\Big)\ ,
\label{2.20}
\end{equation}
with $\{\ ,\, \}$ indicating anticommutator,
and its imaginary, antisymmetric part $\sigma^{(\omega)}$, with components:
\begin{equation}
\sigma^{(\omega)}_{\mu\nu}=-i\lim_{N\to\infty} \omega\Big( \Big[F^{(N)}(x_\mu),\ F^{(N)}(x_\nu)\Big]\Big)\ .
\label{2.21}
\end{equation}
Although this matrix need not be invertible, it is usually called
the {\it symplectic matrix}~\cite{Verbeure-book}.
Indeed, for a non-degenerate $\sigma^{(\omega)}$, the real $n$-dimensional space $\mathcal{X}$  
becomes a symplectic space.%
\footnote{When $\sigma^{(\omega)}$ is not invertible, one can restrict the discussion
to a suitable, physically relevant subspace of $\mathcal{X}$ for which the restricted $\sigma^{(\omega)}$ becomes
non-degenerate ({\it e.g.} see Sect.\ref{section2.5.1} below).}
As such, it supports a bosonic algebra $\mathcal{W}\big(\mathcal{X}, \sigma^{(\omega)}\big)$,
defined as the complex vector space generated by the linear span of operators
$W(\vec{r}\,)$, with $\vec{r}\in \mathbb{R}^n$, obeying the following algebraic relations:
\begin{eqnarray}
\label{2.22}
W(\vec{r_1})\ W(\vec{r_2}) = W(\vec{r_1}+\vec{r_2})\ 
\e^{ -\frac{i}{2} \vec{r_1}\cdot \sigma^{(\omega)}\cdot \vec{r_2} }\ ,\qquad \vec{r_1}, \vec{r_2}\in \mathbb{R}^n\ ,\\
\Big[W(\vec{r}\,)\Big]^\dagger=W(-\vec{r}\,)=\Big[W(\vec{r}\,)\Big]^{-1}\ ,\qquad W({0})={\bf 1}\ .
\label{2.23}
\end{eqnarray}
These relations are just a generalization of the familiar commutation relations of Weyl operators
constructed with single-particle position and momentum operators; for this reason the unitary operators
$W(\vec{r}\,)$ are called (generalized) {\it Weyl operators}, and the algebra 
$\mathcal{W}\big(\mathcal{X}, \sigma^{(\omega)}\big)$ they generate, a (generalized) {\it Weyl algebra}
\cite{Bratteli,Petz}.
As for any operator algebra, a state $\Omega$ on this Weyl algebra is a positive, normalized linear functional
$\Omega:\ \mathcal{W}\big(\mathcal{X}, \sigma^{(\omega)}\big)\to \mathbb{C}$, assigning its mean value
to any element of the algebra. The so-called {\it quasi-free} states $\Omega_\Sigma$
form an important class of such states:
they are characterized by giving a mean value to Weyl operators in Gaussian form \cite{Ferraro, Holevo},
\begin{equation}
\Omega_\Sigma\Big( W(\vec{r}\,) \Big)=\e^{-\frac{1}{2} \vec{r}\cdot\Sigma\cdot \vec{r} }\ ,\qquad \vec{r}\in\mathbb{R}^n\ .
\label{2.24}
\end{equation}
The covariance $\Sigma$ is a positive, symmetric matrix, which, together with the symplectic matrix,
obeys the condition
\begin{equation}
\Sigma + \frac{i}{2}\sigma^{(\omega)}\geq 0\ ,
\label{2.25}
\end{equation}
thus assuring the positivity of $\Omega_\Sigma$. Quasi-free states are {\it regular} states,%
\footnote{ A state $\Omega$ on the Weyl algebra $\mathcal{W}$ is called regular if for any real constant $\alpha$
the map $\alpha \to \Omega\big( W(\alpha\,\vec{r_1}+\vec{r_2})\big)$ is continuous, for all 
$\vec{r_1},\ \vec{r_1}\in \mathbb{R}^n$ \cite{Bratteli}. 
Also {\it irregular} states of Weyl algebras have interesting physical
applications; for a recent account, see \cite{Strocchi4}.}
and as such they admit a representation in terms of Bose fields. 
Let us denote by $\pi_{\Omega_\Sigma}$ the GNS-representation
based on the quasi-free state $\Omega_\Sigma$; then, in this representation, the Weyl operators can be expressed as:
\begin{equation}
\pi_{\Omega_\Sigma}\Big[ W(\vec{r}\,) \Big] = \e^{i \vec{r}\cdot \vec{F}} \ ,
\label{2.26}
\end{equation}
in terms of $n$ (unbounded) Bose operators $F_\mu$, $\mu=1,2,\ldots, n$. They provide an explicit
expression for the associated covariance matrix as their anticommutator:
\begin{equation}
\Sigma_{\mu\nu}=\frac{1}{2} \Omega_\Sigma\Big( \big\{ F_\mu,\ F_\nu\big\} \Big) \ ,
\label{2.27}
\end{equation}
while, thanks to the algebraic relation~(\ref{2.22}), their commutator gives the symplectic matrix:
\begin{equation}
\sigma^{(\omega)}_{\mu\nu}=-i \big[ F_\mu,\ F_\nu\big] \ .
\label{2.28}
\end{equation}

The analogy of the relations (\ref{2.27}) and (\ref{2.28}) with the results
(\ref{2.20}) and (\ref{2.21}) suggests to consider elements in the quasi-local algebra $\mathcal{A}$
obtained by exponentiating the fluctuations $F^{(N)}(x_r)$ in~(\ref{2.17}),
\begin{equation}
W^{(N)}(\vec{r}\,)\equiv \e^{i\vec{r}\cdot\vec{F}^{(N)}(x)}\ ,
\label{2.29}
\end{equation}
and focus on states $\omega$ for which the expectation $\omega\big( W^{(N)}(\vec{r}\,) \big)$
becomes Gaussian in the large $N$ limit. The operators $W^{(N)}(\vec{r}\,)$ will be called {\it Weyl-like}
operators as they behave as true Weyl operators in the thermodynamic limit.
Indeed, let us consider the product of two Weyl-like operators; using the Baker-Campbell-Hausdorff formula,
we can write:
\begin{eqnarray}
\nonumber
&& W^{(N)}(\vec{r_1}\,)\ W^{(N)}(\vec{r_2}\,)=\exp\Bigg\{i F^{(N)}(x_{r_1+r_2})
-\frac{1}{2}\Big[F^{(N)}(x_{r_1}),F^{(N)}(x_{r_2})\Big]\\
\nonumber
&&\hskip 4cm -\frac{i}{12}\Bigg(\Big[F^{(N)}(x_{r_1}),\left[F^{(N)}(x_{r_1}),F^{(N)}(x_{r_2})\right]\Big]\\
&&\hskip 4cm -\Big[F^{(N)}(x_{r_2}),\left[F^{(N)}(x_{r_1}),F^{(N)}(x_{r_2})\right]\Big]\Bigg)+\dots\Bigg\}\ .
\nonumber
\end{eqnarray}
As already seen, in the large $N$ limit, the first commutator on the r.h.s. is proportional to the identity,
while all the additional terms vanish in norm; for instance, one has
\begin{eqnarray}
\nonumber
&&\lim_{N\to\infty}\bigg\|\Big[F^{(N)}(x_{r_1}),\left[F^{(N)}(x_{r_1}),F^{(N)}(x_{r_2})\right]\Big]\bigg\|=\\
&&=\lim_{N\to\infty}\frac{1}{N^{3/2}}
\left\|\sum_{k=1}^N\Big[x_{r_1}^{[k]},\left[x_{r_1}^{[k]},x_{r_2}^{[k]}\right]\Big]\right\|
\le\lim_{N\to\infty}\frac{4}{\sqrt{N}}\|x_{r_1}\|^2\, \|x_{r_2}\|=0\, .
\nonumber
\end{eqnarray}
Therefore, in the thermodynamic limit the Weyl-like operators are seen to obey the following algebraic relations:
\begin{equation}
W^{(N)}(\vec{r_1}\,)\ W^{(N)}(\vec{r_2}\,)\simeq 
W^{(N)}(\vec{r_1} + \vec{r_2}\,)\ \e^{-\frac{1}{2}\big[F^{(N)}(x_{r_1}),F^{(N)}(x_{r_2})\big]}\ ,
\label{2.30}
\end{equation}
which, recalling (\ref{2.21}), reduce to the Weyl relations (\ref{2.22}). In other terms, under suitable
conditions, in the large $N$ limit the operators $W^{(N)}(\vec{r}\,)$ behave as the Weyl operators $W(\vec{r}\,)$
of the algebra $\mathcal{W}\big(\mathcal{X}, \sigma^{(\omega)}\big)$.
The precise way in which this statement should be understood is provided by the following result:

\begin{theorem}
\label{theorem1}
Given the quasi-local algebra $\mathcal{A}$ and the real linear vector space $\mathcal{X}$ as in (\ref{2.16}), and
a clustering state $\omega$ on $\mathcal{A}$, satisfying the conditions:
\begin{eqnarray}
\label{2.31}
&&  1)\ \sum_{k\in\mathbb{N}}\left|\omega\big(x_{r_1}^{[1]}\, x_{r_2}^{[k]}\big)\,
-\,\omega(x_{r_1})\omega(x_{r_2})\right|<\infty\ ,\qquad \vec{r_1},\ \vec{r_2}\in\mathbb{R}^n\\
&&2)\ \lim_{N\to\infty}\omega\Big( \e^{i\vec{r}\cdot\vec{F}^{(N)}(x)}\Big)=
\e^{-\frac{1}{2} \vec{r}\cdot\Sigma^{(\omega)}\cdot \vec{r} }\ ,\qquad \vec{r}\in\mathbb{R}^n\ ,
\label{2.32}
\end{eqnarray}
one can define a Gaussian state $\Omega$ on the Weyl algebra
${\cal W}(\mathcal{X},\sigma^{(\omega)})$ such that, for all $\vec{r}_i\in\mathbb{R}^n$, $i=1,2,\ldots,m$,
\begin{equation}
\hskip -1.5cm \lim_{N\to\infty}\omega\Big( W^{(N)}(\vec{r}_1)\,W^{(N)}(\vec{r}_2)\,\cdots W^{(N)}(\vec{r}_m)\Big)=
\Omega\Big( W(\vec{r}_1)\,W(\vec{r}_2)\,\cdots W(\vec{r}_m)\Big)\ ,
\label{2.33}
\end{equation}
with
\begin{equation}
\lim_{N\to\infty}\omega\Big( W^{(N)}(\vec{r}\,)\Big)=
\e^{-\frac{1}{2} \vec{r}\cdot\Sigma^{(\omega)}\cdot \vec{r}}=
\Omega\Big( W(\vec{r}\,)\Big)\ ,\qquad  \vec{r}\in\mathbb{R}^n\ .
\label{2.34}
\end{equation}
\end{theorem}
\medskip
\noindent
Notice that the Gaussian state $\Omega$ on the algebra $\mathcal{W}(\mathcal{X},\sigma^{(\omega)})$,
with covariance matrix $\Sigma^{(\omega)}$,
is indeed a well defined state. First of all, it is normalized as easily seen by
setting $\vec{r}=0$ in (\ref{2.34}). Further, its positivity is guaranteed
by the positivity of the correlation matrix (\ref{2.18}):
$$
C^{(\omega)}=\Sigma^{(\omega)} + \frac{i}{2} \sigma^{(\omega)}\ge 0\ .
$$
Being Gaussian, the state $\Omega$ gives rise to a regular representation of the Weyl algebra
$\mathcal{W}(\mathcal{X},\sigma^{(\omega)})$, so that one can introduce the Bose fields $F_\mu$
as in (\ref{2.26}) and, through (\ref{2.29}) and (\ref{2.34}), {\it i.e.}
$\lim_{N\to\infty}\omega\big( \e^{i\vec{r}\cdot\vec{F}^{(N)}(x)}\big)=\Omega\big( \e^{i \vec{r}\cdot \vec{F}} \big)$,
identify the large $N$ limit of local fluctuation operators with those Bose fields:
\begin{equation}
\lim_{N\to\infty}F^{(N)}(x_\mu)=F_\mu\ ,\qquad \mu=1,2,\ldots,n\ .
\label{2.35}
\end{equation}
Let us stress that these fields, despite being collective operators, retain
a quantum, non-commutative character. They
describe the behaviour of many-body systems at a level that is half way
between the microscopic world of single-particle observables and the macroscopic realm
of mean-field operators discussed earlier.
In this respect, the large $N$ limit that allows to pass from the exponential
(\ref{2.29}) of the local fluctuations (\ref{2.14})
to the mesoscopic operators belonging to the Weyl algebra $\mathcal{W}(\mathcal{X},\sigma^{(\omega)})$,
as described by the previous Theorem,
can be called the {\it mesoscopic limit}. It can be given a formal definition:

\bigskip\noindent
\textbf{Mesoscopic limit.} {\it 
Given an operator $O^{(N)}$, linear combination of exponential operators $W^{(N)}(\vec{r}\,)$, 
we shall say that it possesses the mesoscopic limit $O$, writing
$$
m-\lim_{N\to\infty} O^{(N)}= O\ ,
$$
if and only if 
\begin{equation}
\hskip -1.5cm \lim_{N\to\infty} \omega\Big( W^{(N)}(\vec{r}_1)\,O^{(N)}\,W^{(N)}(\vec{r}_2)\Big)=
\Omega\Big( W(\vec{r}_1)\,O\,W(\vec{r}_2)\Big)\ ,\quad \forall\, \vec{r}_{1},\ \vec{r}_{2}\in \mathbb{R}^n\ .
\label{2.36}
\end{equation}
} 
\bigskip

\noindent
Note that, by varying $\vec{r}_{1},\ \vec{r}_{2}\in \mathbb{R}^n$, the expectation values of the form
$\Omega\big( W(\vec{r}_1)\,O\,W(\vec{r}_2)\big)$ completely determine any
generic operator $O$ in the Weyl algebra $\mathcal{W}(\mathcal{X},\sigma^{(\omega)})$: essentially,
they represent its corresponding matrix elements.%
\footnote{In more precise mathematical terms, the r.h.s of (\ref{2.36}) corresponds to the
matrix elements of the operator $\pi_\Omega(O)$ with respect to the two vectors
$\pi_\Omega(W(\vec{r}_1) )|\Omega\rangle$, $\pi_\Omega(W(\vec{r}_2)) |\Omega\rangle$ 
in the GNS-representation of the Weyl algebra
$\mathcal{W}(\mathcal{X},\sigma^{(\omega)})$ based on the state $\Omega$ \cite{Bratteli}. Since these vectors are dense in
the corresponding Hilbert space, those matrix elements completely define the operators $O$.}

Similar considerations can be formulated concerning the dynamics of many-body systems at the mesoscopic level.
More precisely,
given a one-parameter family of microscopic dynamical maps $\Phi^{(N)}_t$ on the quasi-local algebra ${\cal A}$,
we will study its action on the Weyl-like operators $W^{(N)}(\vec{r}\,)$,
in the limit of large $N$. In other terms, 
we shall look for the limiting {\it mesoscopic dynamics} $\Phi_t$ acting on the elements $W(\vec{r}\,)$ of the
Weyl algebra $\mathcal{W}(\mathcal{X},\sigma^{(\omega)})$. In line with the previously introduced mesoscopic limit,
to which it reduces for $t=\,0$, we can state the following definition:

\bigskip\noindent
\textbf{Mesoscopic dynamics.} {\it Given a family of one-parameter maps
$\Phi^{(N)}_t\, : {\cal A} \to {\cal A}$, we shall say that it gives the
mesoscopic limit $\Phi_t$ on the Weyl algebra $\mathcal{W}(\mathcal{X},\sigma^{(\omega)})$,
$$
m-\lim_{N\to\infty} \Phi^{(N)}_t = \Phi_t\ ,
$$
if and only if
\begin{equation}
\hskip -1.5 cm
\lim_{N\to\infty} \omega\Big( W^{(N)}(\vec{r}_1)\,\Phi^{(N)}_t\big[W^{(N)}(\vec{r}\,)\big]\, 
W^{(N)}(\vec{r}_2)\Big)=
\Omega\Big( W(\vec{r}_1)\,\Phi_t\big[W(\vec{r}\,)\big]\, W(\vec{r}_2)\Big)\ ,
\label{2.37}
\end{equation}
for all $\vec{r},\ \vec{r}_{1},\ \vec{r}_{2}\in \mathbb{R}^n$.}

\bigskip

\subsection{Spin and oscillator many-body systems}
\label{section2.5}

In order to make more transparent the definitions and results so far presented,
we shall now briefly consider physically relevant models in which the whole treatment can
be made very explicit. 

\subsubsection{Spin chain.}
\label{section2.5.1}

A paradigmatic example of a many-body system, often discussed in the literature, 
is given by a chain of 1/2 spins. The microscopic description of the system
involves three operators $s_1$, $s_2$ and $s_3$, obeying the $su(2)$-algebra commutation relations:
\begin{equation}
\big[ s_j,\ s_k\big]=i \epsilon_{jk\ell}\, s_\ell\ ,\qquad j,k,\ell = 1,2,3\ .
\label{2.38}
\end{equation}
Together with the identity operator $s_0\equiv {\bf 1}/2$, they generate the single-spin
algebra $\mathfrak{a}$, which in this particular case can be identified with $\mathcal{M}_2(\mathbb{C})$,
the set of all $2\times 2$ complex matrices; this algebra is attached to each site of the chain.
The tensor product of single-site algebras from site $p$ to site $q$, $p\leq q$, as in (\ref{2.3}),
forms the local algebras $\mathcal{A}_{[p,q]}$. The union of these local algebras over all possible
finite sets of sites, together with its completion, gives the quasi-local algebra $\mathcal{A}$:
it contains all the observables of the spin chain.

We shall equip $\mathcal{A}$ with a thermal state $\omega_\beta$, at temperature $1/\beta$,
constructed from the tensor product of single-site thermal states:
\begin{equation}
\omega_\beta=\bigotimes_k\ \omega_\beta^{[k]}\ .
\label{2.39}
\end{equation}
At the generic site $k$, the state $\omega_\beta^{[k]}$ is determined by its expectation on the
basis operators:
\begin{eqnarray}
\nonumber
&&\omega_\beta^{[k]}\Big(s_0^{[k]}\Big)=\frac{1}{2}\ ,\quad 
\omega_\beta^{[k]}\Big(s_1^{[k]}\Big)=\omega_\beta^{[k]}\Big(s_2^{[k]}\Big)=\,0\ ,\\
&&\omega_\beta^{[k]}\Big(s_3^{[k]}\Big)=-\frac{\eta}{2}\ ,\qquad \eta\equiv\tanh\left(\frac{\beta\varepsilon}{2}\right)\ .
\label{2.40}
\end{eqnarray}
It can be represented by a Gibbs density matrix $\rho_\beta^{[k]}$ constructed with the site-$k$ Hamiltonian
\begin{equation}
h^{[k]}= \varepsilon\, s_3^{[k]}\ ,
\label{2.41}
\end{equation}
so that for any operator $x^{[k]}\in\mathfrak{a}^{[k]}$:
\begin{equation}
\omega_\beta^{[k]}\Big(x^{[k]}\Big)=\Tr\Big[ \rho_\beta^{[k]}\, x^{[k]}\Big]\ ,\qquad 
\rho_\beta^{[k]}=\frac{\e^{-\beta h^{[k]}}}{2\cosh(\varepsilon\beta/2)}\ .
\label{2.42}
\end{equation}
For a chain containing a finite number $N$ of sites, the state $\omega_\beta$ in (\ref{2.39})
can similarly be represented by a density matrix as:
\begin{equation}
\rho_\beta^{(N)}=\frac{\e^{-\beta \sum_{k=1}^N h^{[k]}}}{\Tr\bigg[\e^{-\beta \sum_{k=1}^N h^{[k]}}\bigg]}\ .
\label{2.43}
\end{equation}
However, this is not longer possible in the thermodynamic limit; indeed, although
$\rho_\beta^{(N)}$ is always normalized for any $N$, it becomes ill-defined in the large $N$ limit,
since it converges (in norm) to zero:
$$
\lim_{N\to\infty} \left\| \rho_\beta^{(N)} \right\| =
\lim_{N\to\infty} \Bigg( \frac{1}{1+\e^{-\beta}} \Bigg)^N =\,0\ .
$$
In other terms, states of infinitely long chains can not in general be represented by density matrices;
on the other hand, the definition in (\ref{2.39}) is perfectly valid in all situations.

Given the single-site spin operators $s_i$ and the state $\omega_\beta$, one can now construct the
corresponding fluctuations as in (\ref{2.14}):
\begin{equation}
F^{(N)}(s_i)\equiv \frac{1}{\sqrt N}\sum_{k=1}^N \Big(s_i^{[k]}-\omega_\beta(s_i){\bf 1}\Big)\ ,\qquad i=1,2,3\ .
\label{2.44}
\end{equation}
From them, the symplectic matrix $\sigma^{(\beta)}$ in (\ref{2.21}) can be easily computed;
taking into account the tensor product structure of the state $\omega_\beta$, it reduces
to the expectation of the commutator of single-site operators:
\begin{equation}
\sigma^{(\beta)}_{jk}= -i\, \omega_\beta\Big( \big[ s_j,\, s_k\big] \Big)\ ,
\label{2.45}
\end{equation}
so that, explicitly:
\begin{equation}
\sigma^{(\beta)}=\frac{\eta}{2}\
\pmatrix{
0 & -1 & 0\cr
1 & 0 & 0\cr
0 & 0 & 0\cr
}
\ .
\label{2.46}
\end{equation}
Recalling (\ref{2.28}), this matrix reproduces the commutators of the Bose operators $F_i$ obtained
as mesoscopic limit of the three fluctuations (\ref{2.44}); as a result, $F_3$ commutes with all
remaining operators and therefore it represents a classical, collective degree of freedom.
On the contrary, the two suitably rescaled operators 
$\hat{P}=\sqrt{2} F_1/\sqrt\eta$ and $\hat{X}=\sqrt{2} F_2/\sqrt\eta$
obey standard canonical commutations: $[\hat{X},\ \hat{P}]=i$, from which standard Weyl operators
$W(\vec{r}\,)\equiv \e^{i(r_1 \hat{P} + r_2 \hat{X})}$ can be defined. The corresponding Weyl algebra
$\mathcal{W}(\sigma^{(\beta)})$ is equipped with a quasi-free state $\Omega_\beta$,
\begin{equation}
\hskip -1cm \lim_{N\to\infty}\omega_\beta\Big( \e^{i\big[ r_1 F^{(N)}(s_1) + r_2 F^{(N)}(s_2) \big]/\sqrt\eta}\Big)=
\e^{-\frac{1}{4}\big[ (r_1^2 +r_2^2)\coth(\beta/2) \big]}=\Omega_\beta\Big( W(\vec{r}\,) \Big)\ ,
\label{2.47}
\end{equation}
which is again a thermal state: it can be represented by a standard Gibbs density matrix:
\begin{equation}
\Omega_\beta\Big( W(\vec{r}\,)\Big)=\frac{\Tr\Big[ \e^{-\beta H}\, W(\vec{r})\Big]}{\Tr\Big[ \e^{-\beta H}\Big]}\ ,
\label{2.48}
\end{equation}
in terms of the free Hamiltonian
\begin{equation}
H=\frac{1}{2}\Big(\hat{X}^2 + \hat{P}^2\Big)\ .
\label{2.49}
\end{equation}

\subsubsection{Harmonic chain.}
\label{section2.5.2}

As a second example of many-body system, let us consider a chain of independent, free harmonic oscillators:
the oscillator attached to site $k$ is described by the position $\hat{x}^{[k]}$ and momentum $\hat{p}^{[k]}$
variables; these operators obey standard canonical commutation relations, 
$[\hat{x}^{[j]},\, \hat{p}^{[k]}]=i\delta_{jk}$, so that
the single-site algebra $\mathfrak{a}$ is now the Heisenberg algebra. The union of all these algebras
for all sites gives the corresponding quasi-local algebra $\mathcal{A}$, that is usually called the
oscillator algebra: elements of this algebra
are polynomials in all variables $(\hat{x}^{[k]},\ \hat{p}^{[k]})$, $k=1,2,\ldots$.

As in the previous example, we shall equip $\mathcal{A}$ with a thermal state $\omega_\beta$, of the form (\ref{2.39}),
with the single-site components $\omega_\beta^{[k]}$ represented by a Gibbs density matrix $\rho_\beta^{[k]}$
as in (\ref{2.42}), where now:
\begin{equation}
\rho^{[k]}_\beta=\frac{e^{-\beta h^{[k]}}}{\tr\big[e^{-\beta h^{[k]}}\big]}\ ,\quad
h^{[k]}=\frac{\varepsilon}{2} \Big[ \big(\hat{x}^{[k]}\big)^2 + \big(\hat{p}^{[k]}\big)^2 \Big]\ ,
\label{2.50}
\end{equation}
with $\varepsilon$ the oscillator frequency, taken for simplicity to be the same for all sites. The state $\omega_\beta$
clearly satisfies both the translation invariance condition (\ref{2.5}) and the clustering
property (\ref{2.7}): in fact, it is a Gaussian state. In order to show this,
one constructs the Weyl operators
\begin{equation}
\widehat{W}({\vec r}\,)=\e^{i \vec r \cdot \vec R}\ ,\qquad \vec r \cdot \vec R \equiv \sum_i r_i\, R_i\ ,
\label{2.51}
\end{equation}
with $\vec{R}$ the vector with components   
$(\hat{x}^{[1]}, \hat{p}^{[1]},\, \hat{x}^{[2]}, \hat{p}^{[2]}\dots)$, and
$\vec r$ a vector of real coefficients.
Although any element of the oscillator algebra 
can be obtained by taking derivatives of $\widehat{W}({\vec r}\,)$ 
with respect to the components of $\vec{r}$, it is
preferable to deal with Weyl operators, since these are bounded operators, unlike
coordinate and momentum operators. Indeed, the oscillator algebra $\mathcal{A}$
should be really identified with the strong-operator closure 
of the Weyl algebra with respect to the so-called GNS-representation
based on the chosen state $\omega_\beta$ (for details, see \cite{Strocchi5,Thirring1,Bratteli}).
In this way, the algebra $\mathcal{A}$ contains only bounded operators; in the following,
when referring to the oscillator algebra, we will always mean the algebra $\mathcal{A}$
constructed in this way.

The expectation of the Weyl operator $\widehat{W}({\vec r}\,)$ is indeed in Gaussian form, 
\begin{equation}
\omega_\beta\Big(\widehat{W}({\vec r}\,)\Big)=\e^{-\frac{1}{2} ({\vec r} \cdot \Sigma \cdot {\vec r})}\ ,
\label{2.52}
\end{equation}
with a covariance matrix $\Sigma$, whose components $[\Sigma]_{ij}$ are defined through the
anticommutator of the different components $R_i$ of $\vec R$:
\begin{equation}
[\Sigma]_{ij}\equiv\frac{1}{2}\, \omega_\beta\Big( \{ R_i,\, R_j\} \Big)=\frac{1}{2\eta}\, [{\bf 1}]_{ij}\ ,
\label{2.53}
\end{equation}
with $\eta$ as in (\ref{2.40}).
Since the covariance matrix is proportional to the unit matrix, 
the state $\omega_\beta$ exhibits no correlations among different oscillators;
the state is therefore completely separable, as shown by its
product form in (\ref{2.39}).

As it will be useful in the following, 
we shall now focus on the following two quadratic elements of the single-site
algebra $\mathfrak{a}$:
\begin{equation}
x_1=\frac{\sqrt{\eta}}{2}(\hat{x}^2-\hat{p}^2)\ , \hskip 2cm   
x_2=\frac{\sqrt{\eta}}{2} \left(\hat{x}  \hat{p}+\hat{p} \hat{x}\right)\ ;
\label{2.54}
\end{equation}
given the real, linear span $\mathcal{X}=\big\{x_r\ \big|\ x_r\equiv\vec{r}\cdot\vec{x}= r_1\, x_1 + r_2\, x_2,
\ \vec{r}\in\mathbb{R}^2\big\}$, let us consider the corresponding fluctuation operators,
defined as in (\ref{2.17}):
\begin{equation}
F^{(N)}(x_r)=r_1\, F^{(N)}(x_1) + r_2\, F^{(N)}(x_2)= \vec{r}\cdot\vec{F}^{(N)}(x)\ .
\label{2.55}
\end{equation}
One easily checks that the large $N$ behaviors of the average of the Weyl-like operator 
obtained by exponentiating these fluctuations, 
$W^{(N)}(\vec{r}\,)\equiv \e^{\vec{r}\cdot\vec{F}^{(N)}(x)}$,
is Gaussian:
\begin{equation}
\lim_{N\to\infty} \omega_\beta\Big( W^{(N)}(\vec{r}\,)\Big)=e^{-\frac{1}{2}\vec{r}\cdot\Sigma^{(\beta)}\cdot \vec{r}}\ ,
\qquad \Sigma^{(\beta)}=\frac{\eta^2+1}{4\eta}\,{\bf 1}_{2}\ ,
\label{2.56}
\end{equation}
where with $\bold{1}_n$ we indicate the unit matrix in $n$-dimension. In addition,
the product of two Weyl-like operators behave as a single one:
\begin{equation}
W^{(N)}(\vec{r}_1)\, W^{(N)}(\vec{r}_2)\sim W^{(N)}(\vec{r}_1+\vec{r}_2)\ e^{-\frac{i}{2}\vec{r}_1\cdot\sigma\cdot\vec{r}_2}\ ,
\label{2.57}
\end{equation}
with a symplectic matrix proportional to the second Pauli matrix $\sigma=i\sigma_2$. This allows defining
collective position $\hat{X}$ and momentum $\hat{P}$ operators,
\begin{equation}
\lim_{N\to\infty}F^{(N)}(x_1)=\hat{X}\ ,\qquad \lim_{N\to\infty}F^{(N)}(x_2)=\hat{P}\ ,
\label{2.58}
\end{equation}
such that $[\hat{X},\ \hat{P}]=i$, and a Gaussian state $\Omega_\beta$ on the corresponding algebra
$\mathcal{W}(\mathcal{X},\sigma)$ of Weyl operators $W(\vec{r}\,)=\e^{r_1 \hat{X} + r_2 \hat{P}}$,
such that
\begin{equation}
\lim_{N\to\infty}\omega_\beta\Big( W^{(N)}(\vec{r}\,)\Big)=
\e^{-\frac{1}{2} \vec{r}\cdot\Sigma^{(\beta)}\cdot \vec{r}}=
\Omega_\beta\Big( W(\vec{r}\,)\Big)\ .
\label{}
\end{equation}
The state $\Omega_\beta$ is again thermal: it can be represented by a single-mode Gibbs density
matrix, in terms of a free oscillator Hamiltonian in the variables $\hat{X}$ and $\hat{P}$.

\section{Quantum fluctuation dynamics}
\label{section3}

In the previous Section we have introduced and studied a class of many-body observables,
the quantum fluctuations, that appear to be the most appropriate for analyzing
system properties at the mesoscopic scale. So far we have devoted our attention to
the ``kinematics'' of such collective observables; in this Section instead 
we shall analyze their dynamical properties. More specifically, we shall study what kind
of dynamics emerges at the mesoscopic level starting from a given microscopic
time-evolution for the elementary constituents of the many-body system.

As remarked in the Introduction, in actual experimental conditions,
many-body systems can hardly be considered isolated
from their surroundings and need to be treated as open quantum systems.
Although the total system composed by the many-body system plus the environment in which
it is immersed is a closed system and as such its time-evolution is unitary, generated by the
total system-environment Hamiltonian, the sub-dynamics of the system alone, obtained by tracing
over the uncontrollable environment degrees of freedom, is in general irreversible and rather complex, showing
dissipative and noisy effects. However, in many physical situations the interaction with the environment
can be considered to be weak, and correlations in the environment to decay fast with respect to the
typical system time-scale;
in such situations, memory effects can be neglected and
the dynamics of the many-body system can be expressed as an effective, reduced dynamics
involving only the system degrees of freedom. It can be described by a family
of one-parameter ($\equiv$ time) maps, obeying the semigroup property, {\it i.e.}
composing only forward in time: they are called ``quantum dynamical semigroups'' \cite{Alicki1}-\cite{Spohn1};
as such, they are generated by master equations that take a specific form, the
so-called Kossakowski-Lindblad form \cite{Kossakowski1}-\cite{Lindblad}.
Such generalized open dynamics have been widely studied and applied to model many dissipative quantum
effects in optical, molecular and atomic physics.

\subsection{Dissipative microscopic dynamics}
\label{section3.1}

Let us then consider a system composed by $N$ particles described by the local algebra 
$\mathcal{A}^{(N)}\subset \mathcal{A}$ whose microscopic, open dynamics is generated 
by master equations of the following, general form:
\begin{equation}
\hskip -1cm
\partial_tX(t)=\mathbb{L}^{(N)}[X(t)]\ ,\qquad \mathbb{L}^{(N)}[X]=\mathbb{H}^{(N)}[X]\,+\,\mathbb{D}^{(N)}[X]\ ,
\quad X\in\mathcal{A}^{(N)}\ ;
\label{3.1}
\end{equation}
the first contribution,
\begin{equation}
\mathbb{H}^{(N)}[X]=i\Big[H^{(N)},\,X\Big]\ ,
\label{3.2}
\end{equation}
is the purely Hamiltonian one, whose generator $H^{(N)}$ can be taken to be the
sum of single-particle Hamiltonians $h^{[k]}=\big(h^{[k]}\big)^\dagger$:
\begin{equation}
H^{(N)}=\sum_{k=1}^{N}\,h^{[k]}\ ,\qquad {H^{(N)}}^\dagger=H^{(N)}\ ,
\label{3.3}
\end{equation}
while the term $\mathbb{D}^{(N)}$ introduces irreversibility and can be cast in the following, generic
Kossakowski-Lindblad form:
\begin{eqnarray}
\nonumber
\hskip -1cm
\mathbb{D}^{(N)}[X]&=&\sum_{k,\ell=1}^{N}J_{k\ell}\sum_{\alpha,\beta=1}^m\,
D_{\alpha\beta}\,\Big(v_\alpha^{[k]}\,X\,(v_{\beta}^{[\ell]})^\dagger
-\frac{1}{2}\left\{v_{\alpha}^{[k]}\,(v_\beta^{[\ell]})^\dagger\,,\,X\,\right\}\Big)\\
&=&\frac{1}{2}\sum_{k,\ell=1}^{N}J_{k\ell}\sum_{\alpha,\beta=1}^m\, D_{\alpha\beta}\,
\Big(v_\alpha^{[k]}\,\left[X\,,\,(v_\beta^{[\ell]})^\dagger\right]\,
+\,\left[v_\alpha^{[k]}\,,\,X\right]\,(v_\beta^{[\ell]})^\dagger\Big)\ ,
\label{3.4}
\end{eqnarray}
with $v_\alpha^{[k]}$ single-particle operators. While the Hamiltonian contribution does not contain
any interaction among the $N$ particles, in the purely dissipative term $\mathbb{D}^{(N)}$
the mixing action of the operators $v_\alpha$ is weighted by the coefficients 
$J_{k\ell}\,D_{\alpha\beta}$, involving in general different particles.
Altogether, they form the Kossakowski matrix $J\otimes D$; in order to ensure the complete positivity%
\footnote{Complete positivity is a stronger condition than just requiring the positivity of the
time-evolution; it needs to be enforced in order to obtain physically meaningful
dynamics in all physical situations. For a complete discussion see \cite{Benatti-rev}.}
of the generated dynamical maps $\displaystyle\Phi_t^{(N)}={\rm e}^{t\mathbb{L}^{(N)}}$,  
both $J$ and $D$ must be positive semi-definite.%
\footnote{The dissipative generator in (\ref{3.4}) is very general and can be obtained
through standard weak-coupling techniques \cite{Alicki1} starting from a microscopic 
system-environment interaction Hamiltonian of the form
$\sum_{k=1}^N\sum_{\alpha,\beta=1}^m v_\alpha^{[k]} \otimes b_\alpha^{[k]}$, where
$b_\alpha^{[k]}$ are suitable hermitian bath operators.}

In order to enforce translation invariance, one attaches the same hamiltonian to each sites $h^{[k]}=h$, 
and further considers different particle couplings $J_{k\ell}$ of the form
\begin{equation}
J_{k\ell}=J(|k-\ell|)\ ,\qquad J_{kk}=J(0)> 0\ .
\label{3.5}
\end{equation}
Furthermore, we shall assume the strength of the mixing terms to be such that:
\begin{equation}
\lim_{N\to\infty}\frac{1}{N}\sum_{k,\ell=1}^{N}|J_{k\ell}|<\infty\ ;
\label{3.5-1}
\end{equation}
recalling the examples presented at the end of the previous Section involving one-dimensional chain systems, 
this condition establishes a fast decay of the strength of the statistical couplings of far separated
sites along the chains, so that the mixing effects due to the presence of the environment are short-range.%
\footnote{The role of long-range interactions will be discussed in the following Section \ref{section4}.}

Notice that the generator $\mathbb{L}^{(N)}$ does not mediate any direct interaction between different particles.
Nevertheless, the dissipative contribution $\mathbb{D}^{(N)}$ accounts for environment induced 
dissipative effects, as it results by rewriting it as the anti-commutator $\{K^{(N)},\ X\}$  
with the pseudo-Hamiltonian $K^{(N)}$,
$$
K^{(N)}=-\frac{1}{2}\sum_{k,\ell=1}^{N}J_{k\ell}\sum_{\alpha,\beta=1}^m\,
D_{\alpha\beta}\ v_{\alpha}^{[k]}\,(v_\beta^{[\ell]})^\dagger\ ,
$$
plus the additional term
$$
\sum_{k,\ell=1}^{N}J_{k\ell}\sum_{\alpha,\beta=1}^m\, D_{\alpha\beta}\ 
v_\alpha^{[k]}\,X\,(v_{\beta}^{[\ell]})^\dagger\ ,
$$
also known as quantum noise. This last piece contributes to statistical mixing:
indeed, by diagonalizing the non-negative matrix $J\otimes D$ and recasting the corresponding  
contribution to $\mathbb{D}^{(N)}$ into the Kraus-Stinespring form $\sum_a L_a\,X\,L_a^\dagger$ 
of completely positive maps, it gives rise to a map transforming pure states into mixed ones.

Finally, we shall further require the time-invariance of the reference microscopic state $\omega$,
\begin{equation}
\omega\bigg(\Phi_t^{(N)}[X]\bigg)=\omega(X) \Leftrightarrow\,\omega\Big(\mathbb{L}^{(N)}[X]\Big)=0\ ,
\label{3.6}
\end{equation}
so that the initial phase of the many-body system is not disrupted by the dynamics \cite{Strocchi1}.
As we shall see in Section \ref{section4}, the release of condition (\ref{3.6}) gives rise to additional issues in the
definition and interpretation of the properties of the fluctuation operators, opening the way 
to the possibility of mesoscopic, non-linear and non-Markovian time-evolutions.

\subsection{Mesoscopic dissipative dynamics}
\label{section3.2}

We shall now study the large $N$ limit of the dynamics generated by (\ref{3.1}) when acting on the 
elements of the fluctuation algebra as introduced in Section \ref{section2}, in order to determine what kind of
time-evolution emerges at the mesoscopic level. When the generator $\mathbb{L}^{(N)}$ in (\ref{3.1})
contains only the Hamiltonian part, without any dissipative contribution, the emerging
mesoscopic dynamics turns out to be unitary and reversible \cite{Goderis1,Verbeure-book}.
When the effects induced by the environment are taken into account, and $\mathbb{D}^{(N)}$ is nonvanishing,
the mesoscopic dynamics that emerges in the limit of large $N$ from the local time-evolution 
$\Phi^{(N)}_t={\rm e}^{t\mathbb{L}^{(N)}}$, ${t\geq0}$, generated by (\ref{3.1})-(\ref{3.4}),
is instead a non-trivial dissipative semigroup $\Phi_t$ of completely positive maps on the algebra of fluctuations.

In order to describe these maps explicitly, let us recall that the fluctuation algebra
is constructed starting from the linear span $\mathcal{X}$ ({\it cf.} (\ref{2.16})) of
a selection of $n$ physically relevant single-particle hermitian operators
$x_\mu$, $\mu=1,2,\ldots, n$; out of them, the fluctuations $F^{(N)}(x_r)=\vec{r}\cdot\vec{F}^{(N)}(x)$
and Weyl-like operators $W^{(N)}(\vec{r}\,)= \e^{\vec{r}\cdot\vec{F}^{(N)}(x)}$ are constructed.
In general, there is no guarantee that the action of the generator $\mathbb{L}^{(N)}$ on 
$F^{(N)}(x_r)=\sum_{\mu=1}^N r_\mu\, F^{(N)}(x_\mu)$ would give a single-particle fluctuation 
still belonging to $\mathcal{X}$. 
In order to recover, out of the action of $\mathbb{L}^{(N)}$,
a mesoscopic dynamics for the Weyl algebra $\mathcal{W}(\mathcal{X},\sigma^{(\omega)})$,
the large $N$ limit of the algebra generated by $W^{(N)}(\vec{r}\,)$, one has to assume the linear span
$\mathcal{X}$ be mapped into itself by the generator $\mathbb{L}^{(N)}$, namely that:
\begin{equation}
\hskip -1cm
\mathbb{L}^{(N)}[x^{[k]}_\mu]=\mathbb{H}^{(N)}[x^{[k]}_\mu]\,+\,\mathbb{D}^{(N)}[x^{[k]}_\mu]
=\sum_{\nu=1}^n \mathcal{L}_{\mu\nu}\,x^{[k]}_\nu\ ,\qquad
\mathcal{L}\equiv\mathcal{H}+\mathcal{D}\ ,
\label{3.7}
\end{equation}
where $\mathcal{H}$ and $\mathcal{D}$ are $n\times n$ coefficient matrices specifying the action
of the Hamiltonian $\mathbb{H}^{(N)}$ and dissipative $\mathbb{D}^{(N)}$ contributions
on $x^{[k]}_\mu$. Given the microscopic dynamics, such assumption is not too restrictive: in general, it can
be satisfied by suitably enlarging the set $\mathcal{X}$ of
physically relevant single-particle operators.

With these assumptions, one can show that the mesoscopic dynamics emerging from the large $N$ limit
of the time evolution $\Phi^{(N)}_t$, as specified by (\ref{2.37}),
is again a dissipative semigroup of maps $\Phi_t$
on the Weyl algebra $\mathcal{W}(\mathcal{X},\sigma^{(\omega)})$, transforming Weyl operators 
into Weyl operators. Maps of this kind are called {\it quasi-free} and their
generic form is as follows \cite{Petz}-\cite{Demoen}:
\begin{equation}
\Phi_t\big[W(\vec{r}\,)\big]=\e^{f_t(\vec{r})}\, W(\vec{r}_t)\ ,
\label{3.8}
\end{equation}
with given time-dependent prefactor and parameters $\vec{r}_t$. In the present case, one finds:
\begin{equation}
\vec{r}_t= \mathcal{M}_t{}^T\cdot \vec{r}\ ,\qquad \mathcal{M}_t=\e^{t {\cal L}}\ ,
\label{3.9}
\end{equation}
where $\mathcal{L}$ is the $n\times n$ matrix  introduced in (\ref{3.7}), while $T$ represents matrix transposition.
Instead, the exponent of the prefactor can be cast in the following form:
\begin{equation}
f_t(\vec{r}_t)=-\frac{1}{2}\, \vec{r}_t\cdot \mathcal{K}_t \cdot \vec{r}_t\ ,\qquad
\mathcal{K}_t=\Sigma^{(\omega)} - \mathcal{M}_t\cdot \Sigma^{(\omega)}\cdot\mathcal{M}_t{}^T \ ,
\label{3.10}
\end{equation}
where $\Sigma^{(\omega)}$ is the covariance matrix defined in (\ref{2.20}). With these definitions, one can
state the following result (whose proof can be found in \cite{Carollo2, Carollo3}):

\begin{theorem}
Given the invariant state $\omega$ on the quasi-local algebra $\mathcal{A}$, 
the real linear vector space $\mathcal{X}$ generated by 
the single-particle operators $x_\mu\in\mathfrak{a}$ and the corresponding 
Weyl-like operators $W^{(N)}(\vec{r}\,)=\e^{i \vec{r}\cdot\vec{F}^{(N)}(x) }$,
evolving in time with the semigroup of maps $\Phi^{(N)}_t\equiv\e^{t\mathbb{L}^{(N)}}$, generated by
$\mathbb{L}^{(N)}$ in (\ref{3.1})-(\ref{3.4}) and leaving $\mathcal{X}$ invariant,
the mesoscopic limit
$$
m-\lim_{N\to\infty}\Phi^{(N)}_t\Big[W^{(N)}(\vec{r}\,)\Big]=\Phi_t\left[W(\vec{r}\,)\right]\ ,
$$
defines a Gaussian quantum dynamical semigroup $\left\{\Phi_t\right\}_{t\ge0}$ on the Weyl algebra 
of fluctuations $\mathcal{W}\left(\mathcal{X},\sigma^{(\omega)}\right)$, 
explicitly given by (\ref{3.8})-(\ref{3.10}).
\end{theorem}

The mesoscopic evolution maps $\Phi_t$ are clearly unital, {\it i.e.} they map the
identity operator into itself, as it follows by letting $\vec{r}=\,0$ in (\ref{3.8}).
In addition, they compose as a semigroup; indeed, for all $s,\ t\geq 0$,
\begin{eqnarray*}
\Phi_s \circ \Phi_t \big[ W(\vec{r}\,)\big]&=&\e^{-\frac{1}{2}\big(\vec{r}\cdot \mathcal{K}_t\cdot\vec{r}
+\vec{r}_t\cdot \mathcal{K}_s\cdot\vec{r}_t\big) }\ W\big((\vec{r}_t)_s\big)\\
&=& \e^{-\frac{1}{2}\big(\vec{r}\cdot \mathcal{K}_t\cdot\vec{r}
+\vec{r}\cdot \big(\mathcal{M}_t\, \mathcal{K}_s\,\mathcal{M}_t^T\big)\cdot\vec{r}\big) }\ W\big(\vec{r}_{t+s}\big)\\
&=&\e^{-\frac{1}{2}\, \vec{r}\cdot \mathcal{K}_{t+s}\cdot\vec{r}}\, W\big(\vec{r}_{t+s}\big)
=\Phi_{t+s}\Big[ W(\vec{r}\,) \Big]\ .
\end{eqnarray*}
Further, the maps  $\Phi_t$ are completely positive, since one can easily check that 
the following condition \cite{Demoen} is satisfied (see Appendix D in \cite{Carollo7}):
\begin{equation}
\Sigma^{(\omega)} +\frac{i}{2}\, \sigma^{(\omega)} \geq
\mathcal{M}_t\cdot \Big( \Sigma^{(\omega)} +\frac{i}{2}\, \sigma^{(\omega)}\Big)\cdot\mathcal{M}_t{}^T\ .
\label{3.11}
\end{equation}
Thanks to the properties of unitality and complete positivity, 
the maps $\Phi_t$ obey Schwartz-positivity:
\begin{equation}
\Phi_t\big[X^\dag X\big]\,\geq\,\Phi_t\big[X^\dag\big]\,\Phi_t\big[X\big]\ .
\label{3.12}
\end{equation}
Using this property and the unitarity of the Weyl operators $W(\vec{r}\,)$, one further finds:
\begin{equation*}
\left|{\rm e}^{f_t(\vec{r})}\right|=\big\|\Phi_t \big[ W(\vec{r}\,)\big]\big\|\leq \|W(\vec{r}\,)\|=1\ .
\end{equation*}
This last result also follows from the positivity of the matrix $\mathcal{K}_t$ in (\ref{3.10}):
this is a direct consequence of the time-invariance of the microscopic state $\omega$ with respect
to the microscopic dissipative dynamics $\Phi_t^{(N)}$ \cite{Carollo3,Carollo7}.
For the same reason, also the mesoscopic Gaussian state $\Omega$ is left invariant by the mesoscopic dynamics
$\Phi_t$; indeed, recalling (\ref{2.34}), one has:
\begin{eqnarray*}
\Omega\left(\Phi_t\left[W(\vec{r}\,)\right]\right)&=&{\rm e}^{f_r(t)}\,\Omega\left(W(\vec{r}_t)\right)=
{\rm e}^{-\frac{1}{2}\, \vec{r}\cdot \mathcal{K}_t\cdot\vec{r}
-\frac{1}{2}\vec{r}_t\cdot \Sigma^{(\omega)}\cdot\vec{r}_t}\\ 
&=&
{\rm e}^{-\frac{1}{2}\, \vec{r}\cdot \mathcal{K}_t\cdot\vec{r}
-\frac{1}{2}\vec{r}\cdot \big(\mathcal{M}_t\, \Sigma^{(\omega)}\, \mathcal{M}_t{}^T\big)\cdot\vec{r}}
={\rm e}^{-\frac{1}{2}\, \vec{r}\cdot \Sigma^{(\omega)}\cdot\vec{r}}
=\Omega\left(W(r)\right)\ .
\end{eqnarray*}

More in general, given any state $\hat\Omega$ on the Weyl algebra $\mathcal{W}\left(\mathcal{X},\sigma^{(\omega)}\right)$,
one defines its time-evolution under $\Phi_t$ according to the dual action: $\hat\Omega\mapsto\hat\Omega\circ \Phi_t$.
For states admitting a representation in terms of density matrices, one can then define 
a dual map $\widetilde{\Phi}_t$ acting on any density matrix
$\rho$ on $\mathcal{W}\left(\mathcal{X},\sigma^{(\omega)}\right)$ by sending it into
$\rho(t)=\widetilde{\Phi}_t[\rho]$, according to the duality relation
\begin{equation}
\tr\Big[\widetilde{\Phi}_t[\rho]\,  W(\vec r\,)\Big]=\tr\Big[\rho\;\Phi_t[W(\vec r\,)]\Big]\ .
\label{3.12-1}
\end{equation}
As already observed, useful states on $\mathcal{W}\left(\mathcal{X},\sigma^{(\omega)}\right)$ 
are Gaussian states $\Omega_\Sigma$,
which are characterized by a Gaussian expectation on Weyl operators
({\it cf.} (\ref{2.24})):
\begin{equation}
\Omega_\Sigma\Big(W(\vec{r}\,)\Big)= \tr\Big[\rho_\Sigma\, W(\vec{r}\,)\Big]=\e^{-\frac{1}{2}(\vec{r}\cdot\Sigma\cdot \vec{r})}\ ,
\label{3.13}
\end{equation}
with
\begin{equation}
[\Sigma]_{\mu\nu}\equiv\frac{1}{2} \tr \Big[ \rho_\Sigma\, \big\{ F_\mu,\, F_\nu\big\} \Big]\ ,\qquad \mu,\ \nu=1,\ldots,n\ ,
\label{3.14}
\end{equation}
$\{F_\mu\}$ being the bosonic operators introduced in (\ref{2.26}), $W(\vec{r}\,)=\e^{i \vec{r}\cdot \vec{F}}$.
These states are completely identified by their covariance matrix $\Sigma$; in particular, as already observed,
positivity of $\rho_\Sigma$ is equivalent to the following
condition \cite{Holevo}:
\begin{equation}
\Sigma+\frac{i}{2}\sigma^{(\omega)}\geq0\ .
\label{3.15}
\end{equation}
One can easily verify that the map $\widetilde{\Phi}_t$ transform Gaussian states into Gaussian states:
\begin{eqnarray}
\nonumber
\tr\Big[\widetilde{\Phi}_t[\rho_\Sigma]\, W(\vec{r}\,)\Big] &=&
{\rm e}^{f_r(t)}\, \tr\Big[\rho_\Sigma\, W(\vec{r}_t\,)\Big]\\
&=&\e^{\left(f_r(t)\,-\,\frac{1}{2}(\vec{r}_t\cdot \Sigma\cdot \vec{r}_t)\right)}=
\tr\Big[\rho_{\Sigma(t)}\, W(\vec{r}\,)\Big]\ ,
\label{3.16}
\end{eqnarray}
with the time-dependent covariance matrix $\Sigma(t)$ explicitly given by:
\begin{equation}
\Sigma(t)=\Sigma^{(\omega)}\,-\,\mathcal{M}_t\cdot\Sigma^{(\omega)}\cdot\mathcal{M}_t{}^T+
\,\mathcal{M}_t\cdot\Sigma\cdot\mathcal{M}_t{}^T\ .
\label{3.17}
\end{equation}
From these results, one recovers the time-invariance of the mesoscopic state $\Omega$,
since starting from the initial covariance $\Sigma\equiv\Sigma^{(\omega)}$, the evolution (\ref{3.17}) gives:
$\Sigma(t)=\Sigma^{(\omega)}$.

\subsection{Mesoscopic entanglement through dissipation}
\label{section3.3}

The presence of an external environment typically leads to decohering and mixing-enhancing phenomena; 
dissipation and noise are common effects observed in quantum systems 
weakly coupled to it \cite{Alicki1}-\cite{Chruscinski1}.
Nevertheless, it has also been shown that suitable environments are capable of creating and enhancing 
quantum correlations among quantum systems immersed in them \cite{Plenio1}-\cite{Benatti5}; 
indeed, entanglement can be generated solely through the mixing structure
of the irreversible dynamics, without any direct interaction between the quantum systems.
This mechanism of environment induced entanglement generation
has been studied for systems made of few qubits 
or oscillator modes \cite{Benatti2}-\cite{Benatti5}; in addition, specific protocols have been proposed
to prepare predefined entangled states via the action of suitably
engineered environments \cite{Kraus}-\cite{Muschik2}. 

Instead, using the just established mesoscopic dynamics on the algebra of fluctuations,
we want now to study the possibility of entanglement generation in many-body systems
through a similar purely noisy mechanism.
More specifically, we shall consider bipartite systems using the chain models presented
in Section \ref{section2}, immersed in a common bath, and show that the emergent dissipative quantum dynamics
at the level of fluctuation observables is capable of generating non-trivial quantum correlations.

\subsubsection{Spin chains.}
\label{section3.3.1}

Let us consider a many-body system composed by two spin-1/2 chains, one next to the other, of the type
already discussed in Section \ref{section2.5.1}, both immersed
in a common thermal bath at temperature $T=1/\beta$. A single site in this double chain system is
composed by the corresponding two sites in the two chains and will be labelled by an integer $k$. 
Following the treatment of
Section \ref{section2}, the tensor product spin algebra 
$\mathfrak{a}=\mathcal{M}_2(\mathbb{C})\otimes \mathcal{M}_2(\mathbb{C})$ 
will be attached to each of these sites; it is generated by the sixteen products 
$s_i\otimes s_j$, $i,j=0,1,2,3$, built with the spin operators 
$s_1$, $s_2$, $s_3$ and $s_0={\bf 1}/2$. Note that
the single-site operators $s_i\otimes s_0$ and $s_0\otimes s_i$, $i=1,2,3$, represent single-spin operators,
pertaining to the first, 
the second of the two chains, respectively.
The tensor product of single-site algebras from site $p$ to site $q$, $p\leq q$, as in (\ref{2.3}),
forms the local algebras $\mathcal{A}_{[p,q]}$; the union of these local algebras over all possible
finite sets of sites, together with its completion, gives the quasi-local algebra $\mathcal{A}$.

We shall equip $\mathcal{A}$ with a thermal state $\omega_\beta$, at the bath temperature $1/\beta$, 
constructed from the tensor product of single-site thermal states as in (\ref{2.39}),
$\omega_\beta=\bigotimes_k\ \omega_\beta^{[k]}$; the only
non vanishing single-site expectations are then:\hfill\break
\vbox{
\begin{eqnarray}
\nonumber
&&\omega_\beta^{[k]}\Big(s_3^{[k]}\otimes {\bf 1}\Big)=
\omega_\beta^{[k]}\Big({\bf 1}\otimes s_3^{[k]}\Big)=-\frac{\eta}{2}\ ,\\
&&\omega_\beta^{[k]}\Big(s_3^{[k]}\otimes s_3^{[k]}\Big)=\frac{\eta^2}{4}\ ,\qquad \eta\equiv\tanh\left(\frac{\beta\varepsilon}{2}\right)\ .
\label{3.18}
\end{eqnarray}
}   
As in (\ref{2.42}), $\omega_\beta^{[k]}$ can be represented by a Gibbs density matrix $\rho_\beta^{[k]}$ 
constructed with the site-$k$ Hamiltonian
\begin{equation}
h^{[k]}= \varepsilon\, \Big( s_3^{[k]}\otimes {\bf 1} + {\bf 1}\otimes s_3^{[k]}\Big)\ ,
\qquad \rho_\beta^{[k]}=\frac{\e^{-\beta h^{[k]}}}{2\cosh(\varepsilon\beta/2)}\ .
\label{3.19}
\end{equation}
Being the product of single-site states, the state $\omega_\beta$ does not support any correlation between
the two spin chains; further, it clearly obeys the clustering condition (\ref{2.7}).

Following the general construction discussed in the previous Section, 
we shall now focus on a subset of all single-particle observables, specifically on:
\begin{eqnarray}
\label{3.20-1}
\hskip -1cm
&&\hskip -1cm x_1=4(s_1\otimes s_0)\ ,\ 
x_2=4(s_2\otimes s_0)\ ,\ x_3=4(s_0\otimes s_1)\ , \ x_4=4(s_0\otimes s_2)\ ,\\
\label{3.20-2}
&&\hskip -1cm x_5=4(s_1\otimes s_3)\ , \ x_6=4(s_2\otimes s_3)\ ,\ x_7=4(s_3\otimes s_1)\ ,\ 
x_8=4(s_3\otimes s_2)\ ,
\end{eqnarray}
and on the real linear span $\mathcal{X}$ generated by them 
(we have introduced suitable factors 4 for later convenience).
Observe that $\omega_\beta(x_\mu)=\,0$, $\mu=1,2\ldots, 8$, and further that the condition (\ref{2.31})
is satisfied, since it simply reduces to $\big|\omega_\beta\big(x_{r_1}\, x_{r_2}\big)\big|<\infty$.

Although there are sixteen single-site observables of the form $s_j\otimes s_k$, $j,k=0,1,2,3$,
it turns out that the set of local fluctuation operators,
\begin{equation}
\label{fluctexpl}
F^{(N)}(x_\mu)=\frac{1}{\sqrt{N}}\sum_{k=1}^{N}\Big(x^{[k]}_\mu-\omega(x_\mu){\bf 1}\Big)
=\frac{1}{\sqrt{N}}\sum_{k=1}^{N} x^{[k]}_\mu\ ,
\label{3.21}
\end{equation}
corresponding to the above subset, gives rise to a set of mesoscopic bosonic operators 
$F_\mu$ whose Weyl algebra commutes with the one generated by the remaining eight elements:
it is then consistent to limit the analysis to the eight single-site operators in (\ref{3.20-1}) and (\ref{3.20-2}).
In addition, note that the couple of operators $x_1$, $x_2$ and $x_3$, $x_4$ refer to observables 
belonging to the first, respectively second spin chain: as we shall see, they provide collective operators 
associated to two different mesoscopic degrees of freedom.

In order to explicitly construct the fluctuation algebra corresponding to the chosen linear span $\mathcal{X}$,
one needs first to compute the correlation matrix $C^{(\beta)}$ as defined in (\ref{2.18}).
Since $\omega_\beta$ is a product state, one simply has:
\begin{equation}
\hskip -1cm C^{(\beta)}_{\mu\nu}=\lim_{N\to\infty} \omega\Big( F^{(N)}(x_\mu)\ F^{(N)}(x_\nu)\Big)
={\rm Tr}\Big[\rho_\beta\, x_\mu\, x_\nu\Big]\ ,
\quad \mu,\nu=1,2,\ldots, 8\ ;
\label{3.22}
\end{equation}
the explicit form of this $8\times 8$ matrix can be expressed as a three-fold tensor
product of $2\times 2$ matrices,
\begin{equation}
C^{(\beta)}= 
\left(\bold{1}_2-\eta\,\sigma_1\right)\otimes\bold{1}_2\otimes\left(\bold{1}_2+\eta\,\sigma_2\right)\ ,
\label{3.23}
\end{equation}
where $\sigma_i$ are standard Pauli matrices, while $\bold{1}_2$ is the unit matrix in two dimensions.
In computing tensor products, we adopt the convention in which 
the entries of a matrix are multiplied by the matrix to its right.
Similarly, one easily obtains the corresponding
covariance matrix,
\begin{equation}
\Sigma^{(\beta)}=(\bold{1}_2-\eta\sigma_1)\otimes \bold{1}_2\otimes \bold{1}_2\ ,
\label{3.24}
\end{equation}
and symplectic matrix,
\begin{equation}
\sigma^{(\beta)}=-2i\eta\, (\bold{1}_2-\eta\sigma_1)\otimes\bold{1}_2\otimes\sigma_2 \ ,
\label{3.25}
\end{equation}
so that: $C^{(\beta)}=\Sigma^{(\beta)} +i \sigma^{(\beta)}/2$. The symplectic matrix gives the commutator
of the Bose operators $F_\mu$, the mesoscopic limit of the fluctuations in (\ref{3.21}):
$[F_\mu,\ F_\nu]=i\sigma^{(\beta)}_{\mu\nu}$.

Let now assume that the interaction of the double chain with the bath in which it is immersed be weak, so that
the effects of the environment can be described by a general master equation of the form (\ref{3.1})-(\ref{3.4}).
For the $N$-site Hamiltonian $H^{(N)}$ we take the sum of $N$ copies of the single-site one in (\ref{3.19}),
$H^{(N)}=\sum_{k=1}^N h^{[k]}$. The dissipative
pieces of the generator is instead constructed using the following single-site operators:
\begin{equation}
\hskip -.5cm
v_1=s_+\otimes s_-\ ,\quad v_2=s_-\otimes s_+\ ,\quad
v_3=2(s_3\otimes s_0)\,,\quad v_4=2(s_0\otimes s_3)\ ,
\label{3.26}
\end{equation}
where $s_\pm=s_1 \pm i s_2$, while for the $4\times 4$ matrix $D$ we take:
\begin{equation}
D= {\bf 1}_2\otimes{\bf 1}_2 + \gamma\, \sigma_1\otimes ({\bf 1}_2 +\sigma_1) 
\ .
\label{3.27}
\end{equation}
The parameter $\gamma$ needs to satisfy the condition $|\gamma|\leq 1/2$
in order for $D$ to be positive semi-definite; it encodes the mixing-enhancing power
of the environment.%
\footnote{More general and involved situations can surely be considered \cite{Carollo2,Carollo3}; 
the simplified model discussed here
results nevertheless quite adequate for showing a general physical phenomenon, namely bath-mediated, mesoscopic
entanglement generation.}
With these choices, the dissipative part $\mathbb{D}^{(N)}$ of the generator
$\mathbb{L}^{(N)}$ can be recast in a double commutator form,
so that one explicitly has:
\begin{eqnarray}\nonumber
\mathbb{L}^{(N)}[X]=&&i\varepsilon\sum_{k=1}^N \Big[ s_3^{[k]}\otimes {\bf 1} + {\bf 1}\otimes s_3^{[k]},\ X\Big]\\ 
&&\hskip 1cm +\frac{1}{2}\sum_{k,\ell=1}^{N}J_{k\ell}\sum_{\alpha,\beta=1}^4
D_{\alpha\beta}\Big[\left[v_\alpha^{[k]},\,X\right],\,
(v_\beta^{[\ell]})^\dagger\Big]\ .
\label{3.28}
\end{eqnarray}
Since operators at different sites commute, the action of this generator on any operator $x_\mu^{[k]}$
at site $k$ simplifies to, recalling (\ref{3.5}):
\begin{eqnarray}\nonumber
\mathbb{L}^{(N)}\Big[x_\mu^{[k]}\Big]=&&
i\varepsilon \Big[ s_3^{[k]}\otimes {\bf 1} + {\bf 1}\otimes s_3^{[k]},\ x_\mu^{[k]}\Big]\\ 
&&\hskip 1cm +\frac{J(0)}{2}\sum_{\alpha,\beta=1}^4
D_{\alpha\beta}\Big[\left[v_\alpha^{[k]},\,x_\mu^{[k]}\right],\,
(v_\beta^{[k]})^\dagger\Big]\ ,
\label{3.29}
\end{eqnarray}
and one can check that the linear span $\mathcal{X}$ is mapped to itself by the action of $\mathbb{L}^{(N)}$;
indeed, one finds: $\mathbb{L}^{(N)}\big[ x_\mu^{[k]}\big]=\sum_{\nu=1}^8 \mathcal{L}_{\mu\nu}\ x_\nu^{[k]}$,
with the $8\times 8$ hermitian matrix $\mathcal{L}$ explicitly given by:
\begin{equation}
\mathcal{L}\equiv\mathcal{H} + \mathcal{D}=-i\varepsilon\, \bold{1}_2\otimes\bold{1}_2\otimes\sigma_2
-J(0)\Big(\bold{1}_8-\gamma\, \sigma_1\otimes\sigma_1\otimes\bold{1}_2\Big)\ .
\label{3.30}
\end{equation}
Via the definitions (\ref{3.9}) and (\ref{3.10}), with $\mathcal{L}$ as in (\ref{3.30}), 
one can now explicitly construct the emergent
mesoscopic dynamics $\Phi_t$ on the Weyl algebra of fluctuations $\mathcal{W}(\mathcal{X},\sigma^{(\beta)})$.
As in the general case treated earlier, also in the present case the mesoscopic
dynamics turns out to be a semigroup of unital, completely positive maps, whose generator is at most
quadratic in the fluctuation operators $F_\mu=\lim_{N\to\infty} F^{(N)}(x_\mu)$. Indeed, one finds that the map 
$W_t(\vec{r}\,) \equiv \Phi_t\big[W(\vec{r}\,)\big] =\e^{f_t(\vec{r})}\, W(\vec{r}_t)$ is generated by a master
equation of the form $\partial_t W_t(\vec{r}\,)=\mathbb{L}\big[ W_t(\vec{r}\,)\big]$, with
\begin{equation}
\hskip-1.5cm
\mathbb{L}[W_t]=\frac{i}{2}\,\sum_{\mu,\nu=1}^8  \mathfrak{H}^{(\beta)}_{\mu\nu}\big[F_\mu F_\nu\,,\,W_t\big]
+\sum_{\mu,\nu=1}^8 \mathfrak{D}^{(\beta)}_{\mu\nu}\left(F_\mu\,W_t\,F_\nu\,-\,\frac{1}{2}\big\{ 
F_\mu F_\nu\,,\,W_t\big\}\right);
\label{3.31}
\end{equation}
in this expression, $\mathfrak{H}^{(\beta)}$ represents a Hermitian $8\times 8$ matrix 
and $\mathfrak{D}^{(\beta)}$ a positive semi-definite 
$8\times 8$ matrix, both expressible in terms of the correlation matrix (\ref{3.23}),
the invertible symplectic matrix (\ref{3.25}) and the matrix in (\ref{3.30}):
\begin{eqnarray}
\nonumber
&&\mathfrak{H}^{(\beta)}=-i(\sigma^{(\beta)})^{-1}\left(\mathcal{L}\,C^{(\beta)}\,-\,C^{(\beta)}\,\mathcal{L}^{T}\right)\,(\sigma^{(\beta)})^{-1}\ ,\\
&&\mathfrak{D}^{(\beta)}=(\sigma^{(\beta)})^{-1}\left(\mathcal{L}\,C^{(\beta)}\,+\,C^{(\beta)}\mathcal{L}^{T}\right)(\sigma^{(\beta)})^{-1}\ .
\label{3.32}
\end{eqnarray}

The Weyl algebraic structure $\mathcal{W}(\mathcal{X},\sigma^{(\beta)})$,
associated with chosen set $\mathcal{X}$ and the microscopic thermal state $\omega_\beta$, can be more
appropriately described in terms of four-mode bosonic annihilation and creation operators
$(a_i,\ a_i^\dagger)$, $i=1,2,3,4$, obeying canonical commutation relations:
\begin{equation}
[a_i,\ a_j^\dagger]=\delta_{ij}\ ,\qquad  [a_i,\ a_j]=[a_i^\dagger,\ a_j^\dagger]=\,0\ .
\label{3.33}
\end{equation}
In fact, one can set:
\begin{equation}
F_\mu=\sum_{i=1}^4 f_\mu{}^i\Big( a_i + a_i^\dagger \Big)
\ ,
\label{3.34}
\end{equation}
with $f_\mu{}^i$ complex coefficients, whose nonvanishing entries are explicitly given by:
\begin{eqnarray}
\nonumber
&&f_1{}^1=if_2{}^1=f_3{}^3=if_4{}^3=\sqrt\eta\ ,\\
\label{3.34-1}
&&f_5{}^1=if_6{}^1=f_7{}^3=if_8{}^3=-\eta^{3/2}\ ,\\
\nonumber
&&f_5{}^2=if_6{}^2=f_7{}^4=if_8{}^4=-\left(\frac{\eta}{1-\eta^2}\right)^{1/2}\ .
\end{eqnarray}
From the first line of (\ref{3.34-1}) one deduces that the creation and annihilation operators
$(a_1,\ a_1^\dagger)$ and $(a_3,\ a_3^\dagger)$, coming from the couples of single-site operators
$x_1$, $x_2$ and $x_3$, $x_4$, refer to the first, respectively the second chain.
In other terms, $(a_1,\ a_1^\dagger)$ and $(a_3,\ a_3^\dagger)$ describe
two independent mesoscopic degrees of freedom emerging from distinct chains. 
Instead, $(a_2,\ a_2^\dagger)$ and $(a_4,\ a_4^\dagger)$ result from combinations of spin operators 
involving both chains at the same time.

The fluctuation algebra $\mathcal{W}(\mathcal{X},\sigma^{(\beta)})$, generated by the Weyl operators
$W(\vec{r}\,)=\e^{i\vec{r}\cdot \vec{F}}$, inherits a quasi-free state $\Omega_\beta$ from the
microscopic state $\omega_\beta$; it is defined by the covariance matrix $\Sigma^{(\beta)}$ in (\ref{3.24}), 
through the following expectation:
\begin{equation}
\Omega_\beta\Big( W(\vec{r}\,)\Big)=\e^{-\frac{1}{2} \vec{r}\cdot\Sigma^{(\beta)}\cdot \vec{r}}\ ,
\qquad  \vec{r}\in\mathbb{R}^8\ .
\label{3.35}
\end{equation}
In the formalism of creation and annihilation operators, the state $\Omega_\beta$ can be represented
by the following density matrix,
\begin{equation}
\label{3.36}
\rho_{\Sigma^{(\beta)}}=\frac{{\rm e}^{-\beta\, H}}{\tr\left({\rm e}^{-\beta\,H}\right)} 
\ ,\qquad H=\varepsilon\sum_{i=1}^4a^\dag_i a_i\ ,
\end{equation}
namely by a Gibbs state at inverse temperature $\beta$  
with respect the quadratic hamiltonian $H$, so that
$\displaystyle \Omega_\beta(W)=\tr(\rho_{\Sigma^{(\beta)}}\,W)$, for any $W\in \mathcal{W}(\mathcal{X},\sigma^{(\beta)})$.
As discussed earlier, coming from a time-invariant microscopic state $\omega_\beta$,
also this mesoscopic state is invariant under the action of the mesoscopic dynamics.

These general results can now be used to analyze the dynamical behaviour of the quantum correlations
between the two chains while following the mesoscopic time evolution $\Phi_t$ and in particular
to study the possibility of bath assisted mesoscopic entanglement generation between the
two spin chains.

By {\it mesoscopic entanglement} we mean the existence of mesoscopic states carrying non-local, 
quantum correlations among the collective operators pertaining to different chains. 
More precisely, we shall focus on the modes $(a_1,\, a_1^\dagger)$ and 
$(a_3,\, a_3^\dagger)$, that, as already observed, 
are collective degrees of freedom attached to the first, second chain, respectively.
In order to have a non-trivial dynamics, as initial state we shall take the time-invariant
mesoscopic thermal state in (\ref{3.36}) further squeezed with a common real parameter $\mathfrak{r}$
along the first and third modes. The resulting state is still uncorrelated, but
its corresponding covariance matrix
$\Sigma_\mathfrak{r}^{(\beta)}$, being $\mathfrak{r}$-dependent, is no longer time-invariant;
rather, it will follow the general evolution given in (\ref{3.17}).

\begin{figure}[t]
\center\includegraphics[scale=0.50]{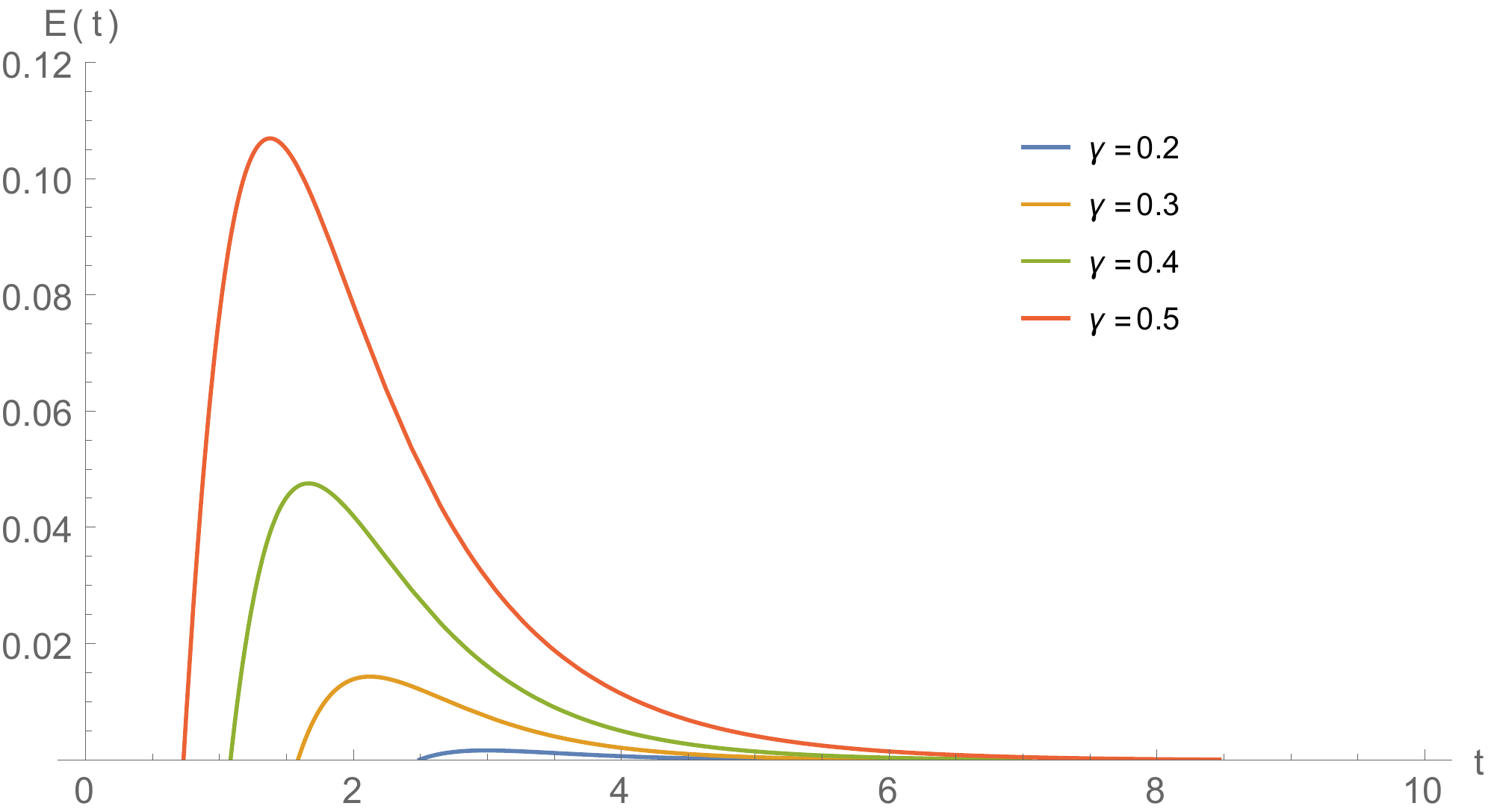}
\caption{\small Spin chain: Behaviour in time of the logarithmic negativity $E(t)$ for different values of the dissipative
coupling $\gamma$, at fixed temperature
$T\equiv1/\beta=1/10$, and squeezing parameter $\mathfrak{r}=1$.}
\label{Fig1}
\end{figure}

\begin{figure}[h!]
\center\includegraphics[scale=0.50]{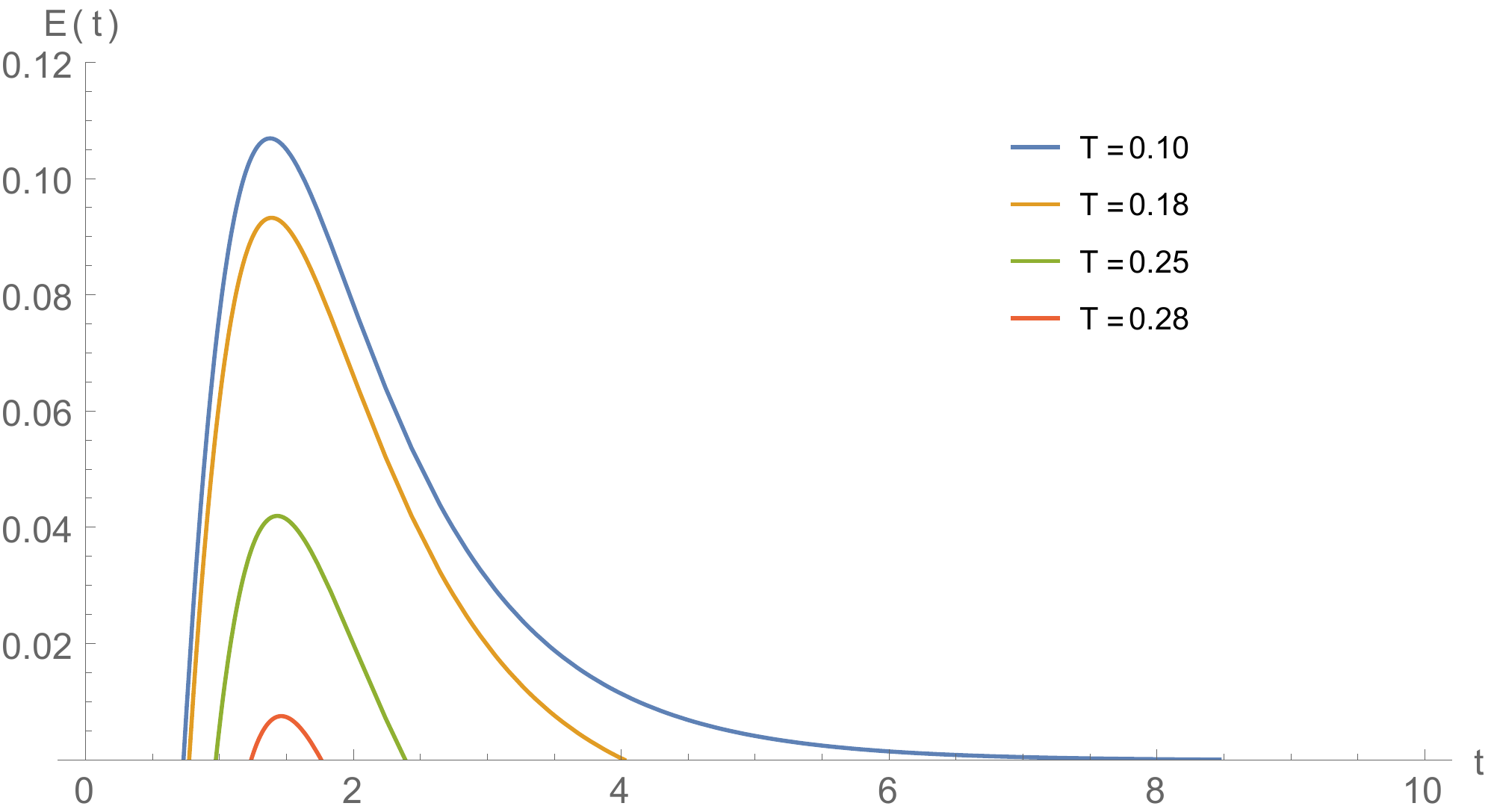}
\caption{\small Spin chain: Behaviour in time of the logarithmic negativity $E(t)$ for different values of the temperature 
$T\equiv 1/\beta$, at fixed dissipative, $\gamma=1/2$, and squeezing, $\mathfrak{r}=1$, parameters.}
\label{Fig2}
\end{figure}

One can now study at any later time $t$ the entanglement content of the reduced, 
two-mode Gaussian state obtained by tracing
over the $(a_2,\ a_2^\dagger)$ and $(a_4,\ a_4^\dagger)$ modes; in practice, one needs to focus on
the reduced covariance matrix, obtained from $\Sigma_\mathfrak{r}^{(\beta)}(t)$ by eliminating
rows and columns referring to the second and fourth mode. Partial transposition
criterion is exhaustive in this case \cite{Simon}, so that entanglement is present between
the remaining first and third collective modes if the smallest symplectic
eigenvalue $\Lambda(t)$ of the partially transposed two-mode, reduced covariant matrix is negative.
Actually, the logarithmic negativity, defined as: 
\begin{equation}
E(t)=\max\Big\{0,\ -\log_2 \Lambda(t)\Big\}\ ,\\
\label{3.37}
\end{equation}
gives a measure of the entanglement content of the state \cite{Souza,Isar}, and it can be analytically
computed for the model under study \cite{Carollo3,Carollo7}. 
One then easily discovers that the dissipative, mesoscopic
dynamics $\Phi_t$ generated by (\ref{3.31}) can indeed produce quantum correlations among
the two initially separable infinite spin chains. As illustrated by the sample behaviour
of $E(t)$ reported in Fig.\ref{Fig1} and Fig.\ref{Fig2}, the amount of created entanglement increases
as the dissipative parameter $\gamma$ gets larger, while it decreases and last for shorter
times as the initial system temperature increases, indicating the existence of a critical
temperature, above which no entanglement is possible.


\subsubsection{Oscillator chains.}
\label{section3.3.2}

In a similar way, one can study the behaviour of a many-body system composed by two infinite
chains of oscillators, {\it i.e.} two copies of the model discussed in Section \ref{section2.5.2}.
As in the previous case, each site $k$ of the double chain consists of a couple of harmonic oscillators,
described by the corresponding position $\hat{x}^{[k]}_\alpha$ and momentum $\hat{p}^{[k]}_\alpha$ operators, the index
$\alpha=1,2$ labelling the two chains; these observables obey a standard Heisenberg algebra,
$\big[ \hat{x}^{[j]}_\alpha,\  \hat{p}^{[k]}_\beta \big]= i\,\delta_{jk}\ \delta_{\alpha\beta}$, and the union
of all these single-site algebras gives the system quasi-local algebra $\mathcal{A}$.
The oscillators are free and therefore their independent microscopic dynamics
is generated by the Hamiltonian:
\begin{equation}
h^{[k]}=\sum_{\alpha=1}^2 h^{[k]}_\alpha\ ,\qquad
h^{[k]}_\alpha = \frac{\varepsilon}{2} \Big[ \big( \hat{x}^{[k]}_\alpha\big)^2 
+ \big( \hat{p}^{[k]}_\alpha\big)^2 \Big]\ ,
\label{3.38}
\end{equation}
with $\varepsilon$ the common oscillator frequency. However, the double chain is assumed immersed 
in a thermal bath, and needs to be treated as an open quantum system. We shall then equip
the system with a thermal state $\omega_\beta$ at the bath temperature $1/\beta$, 
of the product form (\ref{2.39}),
with the single-site components $\omega_\beta^{[k]}$ represented by a Gibbs density matrix 
$\rho_\beta^{[k]}=e^{-\beta h^{[k]}}/\tr\big[e^{-\beta h^{[k]}}\big]$,
with $h^{[k]}$ given by (\ref{3.38}) above.

In order to construct a proper fluctuation algebra for this system, it is convenient to restrict
the discussion to the following single-site, hermitian operators:
\begin{eqnarray}
\nonumber
&& x_1=\frac{\sqrt{\eta}}{2}\Big( (\hat{x}_1)^2- (\hat{p}_1)^2\Big)\ , \hskip 2cm   
x_2=\frac{\sqrt{\eta}}{2} \Big( \hat{x}_1   \hat{p}_1+ \hat{p}_1  \hat{x}_1\Big)\ ,\\
\label{3.39}
&& x_3=\frac{\sqrt{\eta}}{2}\Big( (\hat{x}_2)^2- (\hat{p}_2)^2\Big)\ ,\hskip 2cm  
x_4=\frac{\sqrt{\eta}}{2}\Big(  \hat{x}_2  \hat{p}_2+ \hat{p}_2  \hat{x}_2 \Big)\ ,\\
&& x_5=\sqrt{\frac{2}{\eta}}\, \Big( \hat{x}_1 \hat{x}_2- \hat{p}_1 \hat{p}_2\Big)\ ,\hskip 2.1cm   
x_6 = \sqrt{\frac{2}{\eta}}\, \Big( \hat{x}_1  \hat{p}_2+ \hat{p}_1  \hat{x}_2\Big)\ ,
\nonumber
\end{eqnarray}
with $\eta=\tanh(\beta\varepsilon/2)$, and their corresponding linear span:
\begin{equation}
\mathcal{X}=\Big\{x_r\ \big|\ x_r\equiv\vec{r}\cdot\vec{x}=\sum_{\mu=1}^6 r_\mu\, x_\mu,\ \vec{r}\in\mathbb{R}^6\Big\}\ .
\label{3.40}
\end{equation}
One can then form the quantum fluctuations as in (\ref{2.17}), and study the large $N$ behaviour
of the corresponding Weyl-like operators $W^{(N)}(\vec{r}\,)\equiv \e^{i\vec{r}\cdot\vec{F}^{(N)}(x)}$,
to find:
\begin{equation}
\lim_{N\to\infty} \omega_\beta\Big( W^{(N)}(\vec{r}\,)\Big)=e^{-\frac{1}{2}\vec{r}\cdot\Sigma^{(\beta)}\cdot \vec{r}}\ ,
\qquad \Sigma^{(\beta)}=\frac{\eta^2+1}{4\eta}\,{\bf 1}_{6}\ ,
\label{3.41}
\end{equation}
a generalization of (\ref{2.56}). Together with the covariance matrix $\Sigma^{(\beta)}$, one can also define
a $6\times 6$, antisymmetric, symplectic matrix,
\begin{equation}
\big[\sigma^{(\beta)}\big]_{\mu\nu}=-i\omega_\beta\Big(\big[x_\mu,\, x_\nu\big]\Big)\ ,\qquad
\sigma^{(\beta)}={\bf 1}_3\otimes i\sigma_2\ ,
\label{3.42}
\end{equation}
and thus construct the Weyl algebra of fluctuations $\mathcal{W}(\mathcal{X},\sigma^{(\beta)})$.
Through the mesoscopic limit (\ref{2.36}), the microscopic state $\omega_\beta$
provides a Gaussian state $\Omega_\beta$ on this algebra, so that any of its elements, 
$W(\vec{r}\,)=\e^{i\,\vec{r}\cdot\vec{F}}$, can be represented by means of six collective field operators $F_\mu$, 
obeying canonical commutation relations,
$[F_\mu,\ F_\nu]=i\big[\sigma^{(\beta)}\big]_{\mu\nu}$.

In view of the explicit form (\ref{3.42}) of the symplectic matrix, the components $F_\mu$
can be labelled as
\begin{equation}
\vec{F}=(\hat X_1,\hat P_1,\hat X_2,\hat P_2,\hat X_3,\hat P_3)\ ,
\label{3.43}
\end{equation}
with the $\hat X_i$ position- and $\hat P_i$ momentum-like operators, satisfying
$$
\left[\hat X_i,\hat P_j\right]=i\delta_{i,j}\ ,\qquad i,j=1,2,3\ .
$$
Recalling the definitions (\ref{3.39}), one sees that the couple $\hat X_1$, $\hat P_1$ are operators
pertaining to the first chain of oscillators, while $\hat X_2$, $\hat P_2$ to the second one. On the contrary,
$\hat X_3$, $\hat P_3$ are mixed operators belonging to both chains.
Further, one can show that any other single-site oscillator operator not belonging
to the linear span $\mathcal{X}$ give rise to fluctuation operators that in the large $N$ limit
dynamically decouple from the six in (\ref{3.39}) (see later and \cite{Surace}); 
this is why we can limit the discussion to the chosen set.

Notice that the mesoscopic state $\Omega_\beta$ results separable with respect to the three modes (\ref{3.43}):
its covariance matrix $\Sigma^{(\beta)}$ is diagonal, thus showing neither quantum nor classical
correlations.
Indeed, the state $\Omega_\beta$ can be represented by a density matrix $\rho_\Omega$ 
in product form, $\rho_\Omega=\prod_{i=1}^3 \rho_\Omega^{(i)}$, with
$\rho_\Omega^{(i)}$ standard free oscillator Gaussian states
in the variables $\hat X_i$ and $\hat P_i$.

For a system weakly coupled to the external bath and composed by $N$ sites,
the dynamics can be modelled through the general master equation (\ref{3.1}):
\begin{equation}
\partial_tX(t)=\mathbb{L}^{(N)}[X(t)]\ ,\qquad \mathbb{L}^{(N)}[X]=\mathbb{H}^{(N)}[X]\,+\,\mathbb{D}^{(N)}[X]\ .
\label{3.44}
\end{equation}
The Hamiltonian piece
(\ref{3.2}) involves the total Hamiltonian, $H^{(N)}=\sum_{k=1}^N h^{[k]}$, the sum of $N$
terms of the form (\ref{3.38}). Assuming for simplicity the same bath coupling for all sites, 
the dissipative part of the generator 
$\mathbb{L}^{(N)}$ can be given the following generic structure:
\begin{equation}
\hskip -1cm
\mathbb{D}^{(N)}\left[X\right]\equiv\sum_{k=1}^N \mathbb{D}^{[k]}[X]=\sum_{k=1}^N\sum_{\alpha,\beta=1}^{4} C_{\alpha\beta} 
\left( v^{[k]}_{\alpha} X v^{[k]}_{\beta}-\frac{1}{2}\left\{v^{[k]}_{\alpha}v^{[k]}_{\beta},\, X \right\}\right)\ ,
\label{3.45}
\end{equation}
where $v^{[k]}$ represents the microscopic, site-$k$ operator-valued 
four-vector with components $(\hat{x}_1^{[k]},\ \hat{p}_1^{[k]},\ \hat{x}_2^{[k]},\ \hat{p}_2^{[k]} )$; the $4\times 4$
Kossakowski matrix $C$ with elements $C_{\alpha\beta}$ encodes the bath noisy properties 
and will be taken of the following form:
\begin{equation}
{C}=
\left(
\begin{array}{c|c}
\mathbb{A}  & \mathbb{B} \\ \hline
 \mathbb{B}^\dagger & \mathbb{A}
\end{array}
\right),
\label{3.46}
\end{equation}
with
\begin{equation}
\mathbb{A}=\frac{1+\e^{-\beta\varepsilon}}{2} \Big({\bf 1}_2 -i\eta\, \sigma_2\Big)\ ,
\qquad
\mathbb{B}=\lambda\, \mathbb{A}\ ,
\quad \eta=\tanh(\beta\varepsilon/2)\ .
\label{3.47}
\end{equation}
The first two entries in the four-vector $v^{[k]}$ refer to variables pertaining to the first chain,
while the remaining two to the second chain, so that the diagonal blocks $\mathbb{A}$ of the Kossakowski matrix
describe the evolution of the two chains independently interacting with the same bath; 
in absence of $\mathbb{B}$, the dynamics of the binary system would then be in product form.
Instead, the off-diagonal blocks $\mathbb{B}$ statistically couple the two chains,
and the strength of this coupling is essentially measured by the parameter $\lambda$.%
\footnote{As in the case of the spin chains discussed earlier,
the dissipative generator in (\ref{3.45}) can be obtained through standard weak-coupling techniques
\cite{Alicki1} starting from a microscopic system-environment interaction Hamiltonian of the form
$\sum_{k=1}^N\sum_{\alpha,\beta=1}^{4} v_\alpha^{[k]} \otimes b_\alpha^{[k]}$, where
$b_\alpha^{[k]}$ are suitable hermitian bath operators.}
Further, the condition of complete positivity on the generated dynamics
requires $C$ to be positive semidefinite, which in turn gives
$\lambda^2\leq 1$.

By direct computation, one easily sees that the Kossakowski-Lindblad generator $\mathbb{L}^{(N)}$ above
leaves the linear span $\mathcal{X}$ in  (\ref{3.40}) invariant. Acting on the
fluctuation operators $F^{(N)}(x_\mu)$ of the six single-site variables introduced in (\ref{3.39}),
one explicitly finds:
\begin{equation}
\mathbb{L}^{(N)}\big[\vec{r}\cdot\vec{F}^{(N)}(X)\big]=\vec{r}\cdot\mathcal{L}\cdot \vec{F}^{(N)}(X)\ ,
\label{3.48}
\end{equation}
with:
\begin{equation}
\mathcal{L}=(\e^{-\beta\varepsilon}-1)\, {\bf 1}_6 + 2\varepsilon\, \sigma^{(\beta)} + 
\frac{\lambda(\e^{-\beta\varepsilon}-1)}{\sqrt2}
\left( \begin{array}{lll}
0&{ 0}&{ \bf 1}_2 \\
0&{ 0}&{ \bf 1}_2 \\
{ \bf 1}_2&{ \bf 1}_2&0 \end{array} \right)
\ ,
\label{3.49}
\end{equation}

The master equation (\ref{3.44}), with $\mathbb{D}^{(N)}$ as in (\ref{3.45}), generates a
one-parameter family of transformations $\Phi_t^{(N)}$ mapping Gaussian states 
into Gaussian states \cite{Vanheuverzwijn,Benatti3}, which in the large $N$,
mesoscopic limit gives rise to a quasi-free semigroup of maps $\Phi_t$ on the
Weyl algebra $\mathcal{W}(\mathcal{X},\sigma^{(\beta)})$.
Its explicit form is again as in  (\ref{3.8}), (\ref{3.9}), (\ref{3.10})
with a matrix $\mathcal{L}$ precisely given by (\ref{3.49}). One further checks
that since the starting microscopic thermal state $\omega_\beta$ is left invariant 
by $\Phi_t^{(N)}$, also the mesoscopic Gaussian state $\Omega_\beta$ on
$\mathcal{W}(\mathcal{X},\sigma^{(\beta)})$ is left invariant by the limiting maps $\Phi_t$.

Let us then initially prepare the double chain of oscillators in an uncorrelated state and then investigate whether
the just obtained mesoscopic dynamics is able to generate entanglement between them at the level of collective
observables. More precisely, let us focus on the operators $\hat X_1$, $\hat P_1$ and 
$\hat X_2$, $\hat P_2$, that, as already observed, 
are collective degrees of freedom attached to the first, second chain, respectively. 
One can then study the dynamics of the corresponding reduced, two-mode Gaussian states
by tracing the full three-mode state over the variables $\hat X_3$, $\hat P_3$.

As in the case of spin chains discussed earlier, let us take as initial state the mesoscopic Gaussian
state $\Omega_\beta$, further squeezed with a real parameter $\mathfrak{r}$ along the first two modes.
The entanglement content of the reduced state at any later time $t$ can then be analyzed by looking
at the corresponding logarithmic negativity $E(t)$ as defined in (\ref{3.37}), which also in this
can be analytically computed \cite{Carollo6,Surace}. One easily sees that $E(t)$ become positive in a finite time,
reaching a maximum, whose value increases as the dissipative parameter $\lambda$ gets larger and
the initial bath temperature lowers. 
Since also in this case there are no direct interactions between the bipartite system, as the total Hamiltonian
is that of free, independent oscillators, entanglement between the
two chains is generated at the mesoscopic, collective level by the purely noisy
action of the common environment.

\begin{figure}[h!]
\center\includegraphics[scale=0.35]{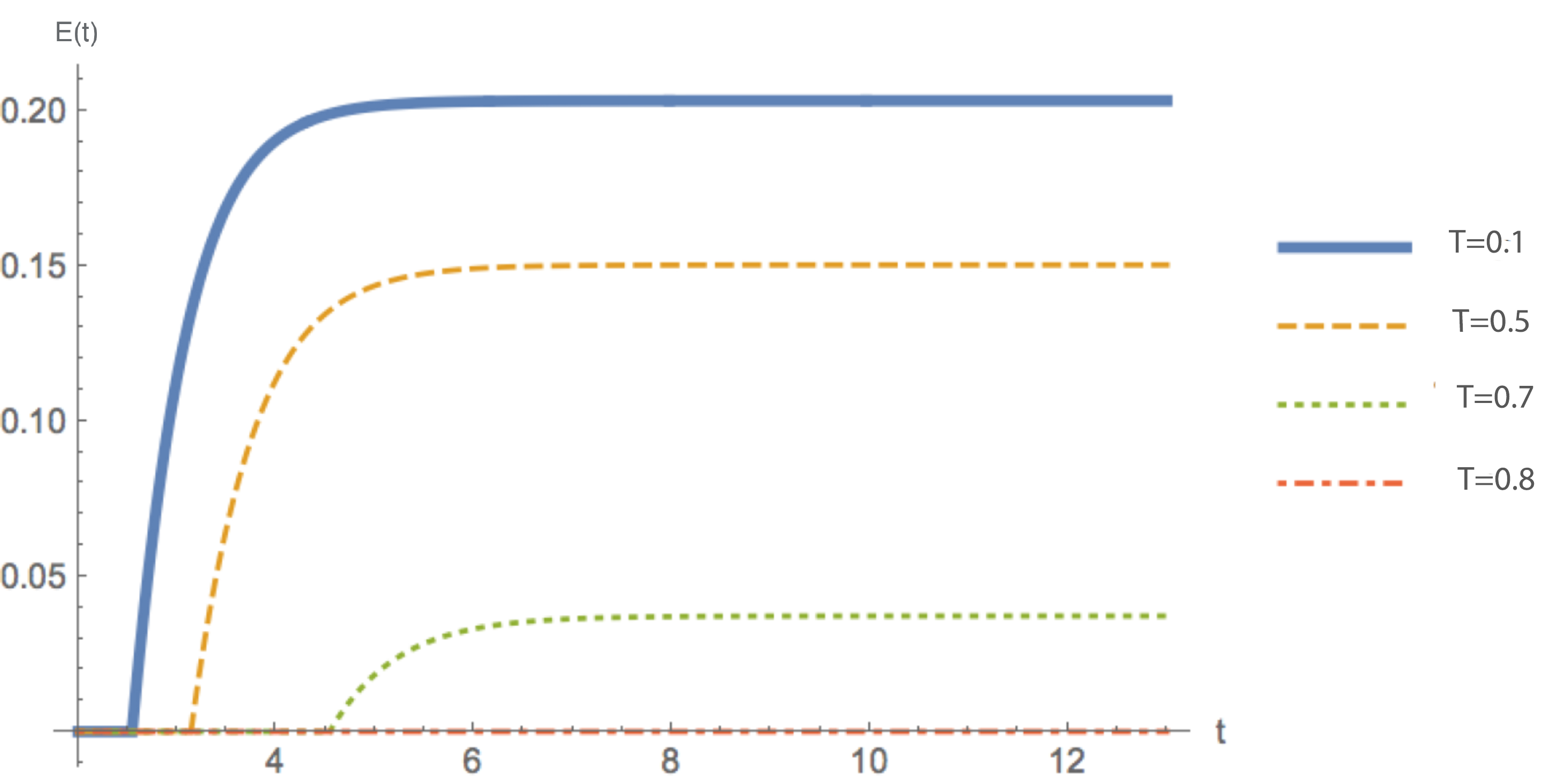}
\caption{\small Oscillator chain: 
Behaviour in time of the logarithmic negativity $E(t)$ for different values of the temperature $T\equiv1/\beta$,
at fixed dissipative, $\lambda=1$, and squeezing, $\mathfrak{r}=1$, parameters. Entanglement rapidly reaches an asymptotic 
nonvanishing value, even at nonzero temperatures.}
\label{Fig3}
\end{figure}

Nevertheless, there is a striking difference between the behaviour of this model and that made of spins.
As observed before, in the case of spin chains entanglement is present only for a finite interval
of time \cite{Eberly}, the larger, the lower the bath temperature is; only in certain specific situations,
involving strictly vanishing temperatures, quantum correlations persist for large times.
Here instead, a non vanishing entanglement can survive for asymptotically long times,
even in presence of a non vanishing initial temperature (see Fig.\ref{Fig3}).
Usually the presence of an environment produces dissipation and noise, ultimately contrasting
the presence of any non-classical correlation; on the contrary, in this case the environment
is able to create and sustain collective quantum correlations among the two chains
for arbitrarily long times and at non-vanishing temperatures. This result clearly
reinforce the possibility of using many-body spintronic and optomechanical systems
in implementing quantum information protocols.

\pagebreak

\section{Long-range interaction systems: mean-field dissipative dynamics}
\label{section4}

In the previous Section, we have discussed the dynamics of quantum fluctuations
in open systems for which the mixing effects due to the presence of the environment are short-range.
Long-range interactions are nevertheless crucial
in explaining coherent phenomena in many-body systems, from phase transitions
to condensation phenomena. Accurate descriptions of these collective effects 
can be obtained through an effective approach, based on the so-called {\it mean-field dynamics},
whose generator scales as the inverse of the number of particles \cite{Alexandrov}-\cite{Hepp}.

As in the previous sections, let us consider a generic many-body system made by a large number $N$
of microscopic constituents, each characterized by the same single-particle algebra of
observables $\mathfrak{a}$, of dimension $d$. It is convenient to fix an orthonormal basis in this algebra,
{\it i.e.} a collection of $d^2$ single-particle, hermitian operators $v_\mu$, $\mu=1,2,\ldots, d^2$,
such that:
\begin{equation}
{\rm Tr}\big(v_\mu\, v_\nu \big)=\, \delta_{\mu\nu}\ .
\label{4.1}
\end{equation}
The unitary, mean-field dynamics for the system is then generated by quadratic interaction Hamiltonians, 
scaling as $1/N$, {\it i.e.} as a mean-field operator ({\it cf.} (\ref{2.8})):
\begin{equation}
H^{(N)}=\frac{1}{N}\sum_{\mu,\nu=1}^{d^2} \mathfrak{h}_{\mu\nu} \sum_{k=1}^N v^{[k]}_\mu \sum_{\ell=1}^N v^{[\ell]}_\nu
=\sum_{\mu,\nu=1}^n \mathfrak{h}_{\mu\nu}\, V^{(N)}_\mu\, V^{(N)}_\nu\ ,
\label{4.2}
\end{equation}
where the hermitian operators
\begin{equation}
V^{(N)}_\mu = \frac{1}{\sqrt N}\sum_{k=1}^N v^{[k]}_\mu\ ,
\label{4.3}
\end{equation}
scale as fluctuations. This Hamiltonian treats each particle on the same footing:
all microscopic constituents of the many-body system interact among themselves with the same strength, 
vanishing as $1/N$ in the large $N$ limit. As such, it can be taken to model long-range interactions 
in many-body systems, providing in many instances a very good description of their dynamical behaviour
in the thermodynamic limit.

For systems in weak interaction with external environments, a common instance in actual experiments,
the reversible, unitary dynamics provided by the previous Hamiltonian should be extended to a dissipative, open
dynamics generated by a suitable Kossakowski-Lindblad operator. The corresponding master equation generating
the time-evolution of any element $X$ in the local algebra
will then take the generic form given in (\ref{3.1}),
\begin{equation}
\hskip -1cm
\partial_t X(t)=\mathbb{L}^{(N)}[X(t)]\ ,\qquad \mathbb{L}^{(N)}[X]=\mathbb{H}^{(N)}[X]\,+\,\mathbb{D}^{(N)}[X]\ ,
\quad X\in\mathcal{A}^{(N)}\ ,
\label{4.4}
\end{equation}
where the first contribution,
\begin{equation}
\mathbb{H}^{(N)}[X]=i\Big[H^{(N)},\,X\Big]\ ,
\label{4.5}
\end{equation}
is the purely Hamiltonian one,
while, in keeping with the structure of (\ref{4.2}), the dissipative term $\mathbb{D}^{(N)}$ can be taken
of the form:
\begin{equation}
\mathbb{D}^{(N)}[X]=\frac{1}{2}\sum_{\mu,\nu=1}^{d^2} C_{\mu\nu}\left(\Big[V^{(N)}_\mu\,,\,X\Big]\,V^{(N)}_\nu\,+\,
V^{(N)}_\mu\,\Big[X\,,\,V^{(N)}_\nu\Big]\right)\ .
\label{4.6}
\end{equation}
In the large $N$ limit, $\mathbb{D}^{(N)}$ scales as $1/N$,
due to the $1/\sqrt{N}$ scaling of the operators $V^{(N)}_\mu$. 
As the Hamiltonian contribution in
(\ref{4.2}) and (\ref{4.5}) models long-range, unitary dynamical effects, similarly
the dissipative piece in (\ref{4.6}) gives rise to long-range mixing effects.
It can can be obtained
through standard weak-coupling techniques \cite{Alicki1} starting from a microscopic 
system-environment interaction Hamiltonian of the form
$\sum_\mu V_\mu^{(N)} \otimes B_\mu$, where
$B_\mu$ are suitable hermitian bath operators;
notice that the $1/\sqrt{N}$ scaling of this interaction Hamiltonian is the same as 
in the Dicke model, used to describe light-matter interaction \cite{Hepp}-\cite{Bagarello}.

As already discussed in the previous Section, the Kossakowski matrix $C_{\mu\nu}$ needs to be non-negative in order
to ensure the complete positivity of the generated dynamical maps 
$\Phi_t^{(N)}={\rm e}^{t\mathbb{L}^{(N)}}$; at the microscopic level,
they will then form a quantum dynamical semigroup of unital maps 
on the local algebra $\mathcal{A}^{(N)}\subset\mathcal{A}$:
\begin{equation}
\Phi_t^{(N)}\circ \Phi_s^{(N)}=\Phi_{t+s}^{(N)}\ ,\quad t,s\geq0\ ,\qquad
\Phi_t^{(N)}[{\bf 1}]={\bf 1}\ .
\label{4.7}
\end{equation}

We shall now study the large $N$ limit of the dynamics generated by (\ref{4.4}) in three different scenarios:
{\it i)} evolution of macroscopic observables, typically the limiting mean-field operators
introduced in Section \ref{section2.3};
{\it ii)} dynamics of microscopic, quasi-local observables, {\it i.e.} operators involving
only a finite number of particles; {\it iii)} emerging mesoscopic dynamics of quantum fluctuations.
These three cases give rise to distinct behaviours,
quite different from the one discussed in the previous Section 
in reference to the master equation (\ref{3.1})-(\ref{3.4}),
whose generator does not scale as a mean-field operator.

\subsection{Dissipative dynamics of macroscopic observables}
\label{section4.1}

We shall start by studying the large $N$ limit of the microscopic dissipative dynamics $\Phi_t^{(N)}$ 
introduced above on the quasi-local algebra $\mathcal{A}$; in other terms, we shall investigate 
the behaviour $\Phi_t^{(N)}[X]$, where $X\in\mathcal{A}$ is either a strictly local element, that is different from the identity matrix, over a fixed, finite number of particles or can be approximated (in norm) by strictly local operators.

As before, we shall consider microscopic states $\omega$ on $\mathcal{A}$ that satisfy the requirements
in (\ref{2.5}) and (\ref{2.7}), {\it i.e.} they are translational invariant and clustering,
but not necessarily invariant under the large $N$ limit of the microscopic dynamics;
in other terms, it might happen that:
\begin{equation}
\lim_{N\to\infty}\omega\bigg(\Phi_t^{(N)}[X]\bigg)\neq\omega(X)\ , \qquad X\in\mathcal{A}\ .
\label{4.8}
\end{equation}
As a result, recalling the discussion of Section \ref{section2.3} according to which mean-field operators 
$\overline{X}^{(N)}=\frac{1}{N}\sum_{k=1}^N x^{[k]}$
become multiple of the identity in the thermodynamic limit,
macroscopic averages associated to these operators might now also change in time.

Let us then focus on the mean-field observables constructed with the single-particle 
basis elements $v_\mu$; their time-evolved averages,
\begin{equation}
\omega_\mu(t):=\lim_{N\to\infty}\omega\left(\Phi_t^{(N)}\left[\frac{1}{N}\sum_{k=1}^{N}v_\mu^{[k]}\right]\right)\ ,
\label{4.9}
\end{equation}
will in general depend on time in the large $N$ limit. In order to write down the equation of motion obeyed
by these macroscopic variables, it is convenient to decompose the coefficients of the mean-field hamiltonian 
in (\ref{4.2}) as $\mathfrak{h}_{\mu\nu}=h_{\mu\nu}+i \kappa_{\mu\nu}$, 
with the real part $h$ and the imaginary one $\kappa$ satisfying the relations 
\begin{equation}
h_{\mu\nu}=h_{\nu\mu}\ ,\qquad \kappa_{\mu\nu}=-\kappa_{\nu\mu}\ .
\label{4.10}
\end{equation}
Similarly, the Kossakowski matrix $C=[C_{\mu\nu}]$ can be decomposed in its self-adjoint symmetric 
and anti-symmetric components as
\begin{equation}
A:=\frac{C+C^{T}}{2}\ ,\quad B:=\frac{C-C^{T}}{2}\ ,\qquad A_{\mu\nu}=A_{\nu\mu}\ ,\ \
B_{\mu\nu}=-B_{\nu\mu}\ ,
\label{4.11}
\end{equation}
where $T$ denotes matrix transposition. Then, the generator $\mathbb{L}^{(N)}$ in (\ref{4.4})
can be rewritten as:
\begin{eqnarray}
\label{4.12-1}
&&\hskip -1cm\mathbb{L}^{(N)}[X]=\mathbb{A}^{(N)}[X]+\mathbb{B}^{(N)}[X]\ ,\\
\label{4.12-2}
&&\hskip -1cm\mathbb{A}^{(N)}[X]=\frac{1}{2}\sum_{\mu,\nu=1}^{d^2}\widetilde{A}_{\mu\nu}\,
\left[\Big[V^{(N)}_\mu\,,\,X\Big]\,,\,V^{(N)}_\nu\right]\ ,\quad
\widetilde{A}_{\mu\nu}:=A_{\mu\nu}\,-\,2\kappa_{\mu\nu}\ ,\\
\label{4.12-3}
&&\hskip -1cm\mathbb{B}^{(N)}[X]:=\frac{1}{2}\sum_{\mu,\nu=1}^{d^2}\widetilde{B}_{\mu\nu}\,
\left\{\Big[V^{(N)}_\mu\,,\,X\Big]\,,\,V^{(N)}_\nu\right\}\ ,\quad
\widetilde{B}_{\mu\nu}:=B_{\mu\nu}\,+\,2i\,h_{\mu\nu}\ .
\end{eqnarray}

By taking the time derivative of (\ref{4.9}) and using the above decomposition for the generator
$\mathbb{L}^{(N)}$, one can deduce that the macroscopic averages $\omega_\mu(t)$
obey the following non-linear equations \cite{Carollo5}:
\begin{equation}
\frac{{\rm d}}{{\rm d}t}\omega_\mu(t)=i\sum_{\alpha,\beta,\gamma=1}^{d^2} 
f_{\alpha\mu}{}^\gamma\, \widetilde{B}_{\alpha\beta}\ \omega_\beta(t)\, \omega_\gamma(t)\ ,\quad \mu=1,2,\cdots,d^2\ .
\label{4.13}
\end{equation}
where $f_{\alpha\beta}{}^\gamma$ are the real structure constant for the basis elements $v_\alpha$
of the single-particle algebra $\mathfrak{a}$,
$[v_\alpha,\ v_\beta]=i\sum_{\gamma=1}^{d^2} f_{\alpha\beta}{}^\gamma\ v_\gamma$.

For later convenience, it is useful to recast this evolution equation in a compact, matrix form;
denoting by $\vec{\omega}_t$ the vector with components $\omega_\mu(t)$, (\ref{4.13}) can be rewritten as
\begin{equation}
\frac{{\rm d}}{{\rm d}t}\vec{\omega}_t=D^{(\vec{\omega}_t)}\cdot\vec{\omega}_t\ ,\qquad
D_{\mu\gamma}^{(\vec{\omega}_t)}=
i\sum_{\alpha,\beta=1}^{d^2} f_{\alpha\mu}{}^\gamma\, \widetilde{B}_{\alpha\beta}\ \omega_\beta(t)\ ,
\label{4.14}
\end{equation}
where the matrix $D^{(\vec{\omega}_t)}$ depends implicitly on time through the time-evolution: $\vec{\omega}\mapsto\vec{\omega}_t$.
Since $\widetilde B$ changes sign under conjugation, 
the matrix $D^{(\vec{\omega}_t)}$ is real; further, it is antihermitian:
$(D^{(\vec{\omega}_t)})^\dagger=-D^{(\vec{\omega}_t)}$.

The non-linear equations (\ref{4.14}), with initial condition $\vec{\omega}_{t=0}=\vec{\omega}$, 
are formally solved by the matrix expression:
\begin{equation}
\vec{\omega}_t=M_t^{(\vec{\omega})}\cdot \vec{\omega}\ ,\quad 
M_t^{(\vec{\omega})}\equiv\mathbb{T}{\rm e}^{\int_0^t{\rm d}s\,D^{(\vec{\omega}_s)}}\ ,
\label{4.15}
\end{equation}
where $\mathbb{T}$ denotes time-ordering; the dependence of the $d^2\times d^2$ matrix $M_t^{(\vec{\omega})}$ 
on the time-evolution $\vec{\omega}\mapsto\vec{\omega}_t$ embodies the non-linearity of the dynamics.
Despite the time-ordering, since there is no explicit time-dependence in the equations (\ref{4.13}), 
the time-evolution of the macroscopic averages composes as a semigroup, 
\begin{equation}
\vec{\omega}\mapsto\vec{\omega}_s\mapsto(\vec{\omega}_s)_t=\vec{\omega}_{s+t}\qquad \forall\ s,t\geq 0\ .
\label{4.16}
\end{equation}
In addition, since the matrix $D^{(\vec{\omega}_t)}$ is antisymmetric and the macroscopic averages are real, 
the quantity $\sum_{\alpha=1}^{d^2}\omega^2_\alpha(t)$ is a constant of motion.

\subsubsection{Spin chain: macroscopic observables}
\label{section4.1.1}

As a specific example, let us consider the many-body system 
introduced in Section \ref{section2.5.1}, given by a chain of 1/2 spins.
In this case, the single-particle algebra $\mathfrak{a}$ coincides with $\mathcal{M}_2(\mathbb{C})$, 
the set of $2\times 2$ complex matrices. A basis in this algebra is given by the three
spin operators $s_1$, $s_2$ and $s_3$ obeying the $su(2)$-algebra commutation relations,
$[s_j,\ s_k]=i\epsilon_{jk\ell}\, s_\ell$, together with the identity $s_0={\bf 1}/2$.

For sake of simplicity, let us consider a purely dissipative mean-field dynamics, generated by 
$\mathbb{L}^{(N)}=\mathbb{D}^{(N)}$, with $\mathbb{D}^{(N)}$ as in (\ref{4.6}),
with:
\begin{equation}
V^{(N)}_\mu = \frac{1}{\sqrt N}\sum_{k=1}^N s^{[k]}_\mu\ , \quad \mu=0,1,2,3\ .
\label{4.17}
\end{equation}
By choosing an environment giving a Kossakowski matrix of the form:
\begin{equation}
C=\pmatrix{
1&i&0\cr
-i&1&0\cr
0&0&0\cr
}\ ,
\label{4.18}
\end{equation}
the generator $\mathbb{D}^{(N)}$ in (\ref{4.6}) can be conveniently recast in the following compact form:
\begin{equation}
\hskip -1.5cm
\mathbb{D}^{(N)}[X]=V_+^{(N)}\, X\, V_-^{(N)}-
\frac{1}{2}\Big\{V^{(N)}_+\,V^{(N)}_-\,,\,X\Big\}\ ,\qquad V^{(N)}_\pm=V^{(N)}_1\pm i\,V^{(N)}_2\ .
\label{4.19}
\end{equation}
The symmetric and anti-symmetric components of $C$ are then given by
\begin{equation}
A=\pmatrix{
1&0&0\cr
0&1&0\cr
0&0&0\cr}
\qquad
B=\pmatrix{
0&i&0\cr
-i&0&0\cr
0&0&0\cr}
\ .
\label{4.20}
\end{equation}

Taking for the microscopic state $\omega$ the thermal state
introduced in Section \ref{section2.5.1},
the only non-trivial macroscopic averages $\omega_\mu(t)$ in (\ref{4.9}) are 
$\omega_{1,2,3}(t)$, while $\omega_0(t)=1/2$ for all $t\geq 0$.
Therefore, one can limit the discussion to the vector $\vec{\omega}_t=\left(\omega_1(t),\omega_2(t),\omega_3(t)\right)$;
since $\|s_\mu\|\leq 1/2$, its components belong to the interval $[-1/2\,,\,1/2]$. 
In the present case, the equations (\ref{4.13}) simply become:
\begin{eqnarray}
\nonumber
&&\frac{{\rm d}}{{\rm d}t}\omega_1(t)=-\omega_1(t)\,\omega_3(t)\ ,\\
\label{4.21}
&&\frac{{\rm d}}{{\rm d}t}\omega_2(t)=-\omega_2(t)\,\omega_3(t)\ ,\\
\nonumber
&&\frac{{\rm d}}{{\rm d}t}\omega_3(t)=\omega^2_1(t)\,+\,\omega^2_2(t)\ ,
\end{eqnarray}
corresponding to the following matrix $D^{(\vec{\omega}_t)}$ as defined in (\ref{4.14}):
\begin{equation}
D^{(\vec{\omega}_t)}=
\pmatrix{
0&0&\omega_1(t)\cr
0&0&\omega_2(t)\cr
-\omega_1(t)&-\omega_2(t)&0
}\ .
\label{4.22}
\end{equation}
As already observed, the norm $\xi\equiv[(\omega_1(t))^2 + (\omega_2(t))^2 + (\omega_3(t))^2]^{1/2}$ 
of $\vec{\omega}_t$ is a constant of motion,
so that the third equation can be readily solved:
\begin{equation}
\omega_3(t)=\xi\tanh\left(\xi(t+c)\right)\ ,
\label{4.23}
\end{equation}
where the constant $c$ is related to the initial condition: 
\hbox{$\omega_3(0)=\xi\tanh\left(\xi c\right)$}.
Inserting this result in the remaining two equations, one further gets:
\begin{equation}
\omega_1(t)=\,\frac{\cosh(c\,\xi)}{\cosh(\xi(t+c))}\,\omega_1\ ,\quad
\omega_2(t)=\,\frac{\cosh(c\,\xi)}{\cosh(\xi(t+c))}\,\omega_2\ ,
\label{4.24}
\end{equation}
where $\omega_{1,2}\equiv\omega_{1,2}(0)$. Notice that the only time-invariant solution
of the equations (\ref{4.21}) is given by: $\omega_1=\omega_2=\,0$ and $\omega_3=\xi$,
which is a stable solution for $\xi\geq 0$; in this case, starting from any initial
triple $(\omega_1, \omega_2, \omega_3)$, one always converges to $(0,0,\xi)$ in the long time limit.

\subsection{Dissipative dynamics of microscopic observables: emergent unitary dynamics}
\label{section4.2}

Having determined the dynamics of the basic macroscopic averages,
we can now focus on the large $N$ time evolution of strictly local observables.
In order to get a hint on the limiting dynamics, let us consider a
single-particle operator $x^{[k]}$ as in (\ref{2.4}), and the action
of the generator (\ref{4.12-1}) on it.
Since operators at different sites commute, the double commutator in the first contribution (\ref{4.12-2}) 
to $\mathbb{L}^{(N)}$ yields
\begin{equation}
\mathbb{A}^{(N)}\big[x^{[k]}\big]=\frac{1}{2N}\sum_{\mu,\nu=1}^{d^2}\widetilde A_{\mu\nu}\Big[\left[v_\mu^{[k]},x^{[k]}\right],v_\nu^{[k]}\Big]\ .
\label{4.25}
\end{equation} 
The norm of $\mathbb{A}^{(N)}\big[x^{[k]}\big]$ vanishes as $N\to\infty$, since the double sum contains
a finite number of contributions, each of them norm bounded. On the other hand, the second contribution
(\ref{4.12-3}) to $\mathbb{L}^{(N)}$ gives:
\begin{equation}
\mathbb{B}^{(N)}\big[x^{[k]}\big]=\frac{1}{2N}\sum_{\ell=1}^N \sum_{\mu,\nu=1}^{d^2}\widetilde B_{\mu\nu}
\Big\{\left[v_\mu^{[k]},x^{[k]}\right],v_\nu^{[\ell]}\Big\}\ .
\label{4.26}
\end{equation}
As discussed in Section \ref{section2.3}, for any clustering state $\omega$, the mean-field observable
$\frac{1}{N}\sum_{\ell=1}^N v_\nu^{[\ell]}$ tends in the large $N$ limit to a scalar quantity, given by
$\omega_\nu\equiv\omega(v_\nu)$.
As a consequence, in the limit, the contribution (\ref{4.26}) becomes a commutator with 
a state-dependent Hamiltonian:
\begin{equation}
\lim_{N\to\infty}\mathbb{B}^{(N)}\big[x^{[k]}\big]=i\left[H^{[k]}_{\vec\omega}\,,\,x^{[k]}\right]\ ,\qquad 
H^{[k]}_{\vec\omega}=-i\sum_{\mu,\nu=1}^3 \widetilde B_{\mu\nu}\,\omega_\nu\,v_\mu^{[k]}\ .
\label{4.27}
\end{equation}
Since $(\widetilde B_{\mu\nu})^*=-\widetilde B_{\mu\nu}$, while the expectations $\omega_\nu$ are real,
it follows that $H^{[k]}_\omega$ is hermitian, as $(v_\nu^{[k]})^\dag=v_\nu^{[k]}$.

However, this result is not sufficient for determining the correct dynamical equation for $x^{[k]}$;
indeed, according to (\ref{4.4}), one should
analyze the action of the generator $\mathbb{L}^{(N)}$ on the time-evolved $x^{[k]}$ at the generic time $t$,
{\it i.e.} on $\Phi_t^{(N)}[x^{[k]}]\equiv{\rm e}^{t\mathbb{L}^{(N)}}[x^{[k]}]$.
Further, one should keep in mind that, as previously discussed,
the state $\omega$ might not be time-invariant in the large $N$ limit: averages of mean-field observables
will in general depend on time. Recalling (\ref{4.9}), this suggests that the Hamiltonian in (\ref{4.27})
should be substituted by a time-dependent one, with $\vec\omega$ replaced by $\vec\omega_t$.
Explicit computation (see \cite{Carollo5}) indeed provides
the expected result: 
$H^{[k]}_{\vec{\omega}_t}=-i\sum_{\mu,\nu=1}^3 \widetilde B_{\mu\nu}\,\omega_\nu(t)\,v_\mu^{[k]}$.

In addition, taking an arbitrary initial time $t_0$, not necessarily $t_0=\,0$ as so far implicitly understood,
the emergent dynamics on strictly local observables $x^{[k]}$ will be the result
of the large $N$ limit of the microscopic dynamical map $\Phi_{t-t_0}^{(N)}={\rm e}^{(t-t_0)\mathbb{L}^{(N)}}$;
as such, it is generated by an Hamiltonian of the form
\begin{equation}
H^{[k]}_{\vec{\omega}_{t-t_0}}=-i\sum_{\mu,\nu=1}^3 \widetilde B_{\mu\nu}\,\omega_\nu(t-t_0)\,v_\mu^{[k]}\ ,
\label{4.28}
\end{equation}
which now explicitly depends on the initial time $t_0$, besides the running time $t$.
In other terms, in the large $N$ limit, the irreversible, dissipative semigroup of maps $\Phi_{t-t_0}^{(N)}$
when acting on local observables gives rise, rather surprisingly, to a family of unitary maps 
$\alpha_{t-t_0}=\lim_{N\to\infty}\Phi_{t-t_0}^{(N)}$.
Further, these automorphisms do not satisfy the microscopic composition law (\ref{4.7}),
nor the one typical of two-parameter semigroups,
$\gamma_{t,t_0}=\gamma_{t,s}\circ\gamma_{s,t_0}$, for any $t_0\leq s\leq t$,
due to the explicit initial-time dependence of their generator. The unitary maps $\alpha_{t-t_0}$
generated by the Hamiltonian in (\ref{4.28}) thus provide an instance of non-Markovian evolution 
as defined in \cite{Chruscinski2}.

The extension of these results to any quasi-local operator $X\in\mathcal{A}$ is given by the following
theorem \cite{Carollo5}.

\begin{theorem}
\label{theorem3}
Given a translation-invariant state $\omega$ on the quasi-local algebra
$\mathcal{A}$ satisfying the $L_1$-clustering prooperty (\ref{2.12}), 
in the large $N$ limit the local dissipative generator $\mathbb{L}^{(N)}$ 
in (\ref{4.12-1})-(\ref{4.12-3}) defines on $\mathcal{A}$ a one-parameter family of automorphisms 
that depend on the state $\omega$ and are such that, for any initial time $t_0\geq 0$, 
\begin{equation}
\lim_{N\to\infty}\omega\left(A\,\Phi^{(N)}_{t-t_0}[X]\,B\right)=\omega\Big(A\,\alpha_{t-t_0}[X]\,B\Big)\ ,
\label{4.29}
\end{equation}
for all $A,B,X\in\mathcal{A}$. If $X$ has finite support, {\it i.e.} it involves only a finite number $S$
of particles, then
\begin{equation} 
\alpha_{t-t_0}[X]=\big(U^{(S)}_{t-t_0}\big)^\dag\,X\,U^{(S)}_{t-t_0}\ ,\quad
U^{(S)}_{t-t_0}=\mathbb{T}{\rm e}^{-i\int_0^{t-t_0}{\rm d}u\, H^{(S)}_{\vec{\omega}_u}}\ ,
\label{4.30}
\end{equation}
with an explicitly time-dependent Hamiltonian:
\begin{equation}
\label{auto3}
H^{(S)}_{\vec{\omega}_t}=-i\,\sum_{k=1}^{S}\,\sum_{\mu,\nu=1}^{d^2}\widetilde{B}_{\mu\nu}\,\omega_\nu(t)\,v_\mu^{(k)} \ .
\label{4.31}
\end{equation}
\end{theorem}
Let us remark that the convergence of the mean-field dissipative 
dynamics $\Phi^{(N)}_{t-t_0}$ to the automorphism $\alpha_{t-t_0}$ 
of $\mathcal{A}$ should be intended in the weak-operator topology associated 
with the GNS-representation based on the state $\omega$.
Further, notice that the automorphisms $\alpha_t$, being the limit of the semigroup $\Phi^{(N)}_t$,
have a meaning only for $t\geq0$; in other terms, 
although the inverted automorphisms $(\alpha_t)^{-1}=\alpha_{-t}$ surely exist, 
they have no physical relevance, since they can not arise from the underlying 
non-invertible microscopic dynamics. Finally, as previously explained, the
automorphisms $\alpha_{t-t_0}$ represent an example of non-Markovian time evolution;
nevertheless, when $\lim_{N\to\infty}\omega\circ\Phi_t^{(N)}$ 
provides a time-invariant state on the quasi-local algebra $\mathcal{A}$,
then one recovers for $\alpha_t$ the one-parameter semigroup composition law as in (\ref{4.7}).

\subsubsection{Spin chain: quasi-local observables}
\label{section4.2.1}

Let us now reconsider the example of a spin-1/2 chain presented in Section \ref{section4.1.1}.
From the specifications and results collected there, one can immediately 
write down the explicit expression of the Hamiltonian in (\ref{4.31})
involving the first $S$ sites of the chain:
\begin{equation}
H^{(S)}_{\vec{\omega}_t}=\sum_{k=1}^{S}\left(\omega_1(t)s_2^{[k]}-\omega_2(t)s_1^{[k]}\right)\ .
\label{4.32}
\end{equation}
Observe that this Hamiltonian commute with itself at different times,
$[H^{(S)}_{\vec{\omega}_{t_1}}\,,\,H^{(S)}_{\vec{\omega}_{t_2}}]=0$, for all $t_{1}$, $t_2$, 
so that the time-ordering in the definition of the unitary operator
implementing the finite time evolution in (\ref{4.30}) is irrelevant.
Then, starting the dynamics at $t_0=\,0$, one easily finds:
\begin{equation}
U^{(S)}_t = {\rm e}^{-i\int_0^t{\rm d}u\, H^{(S)}_{\vec{\omega}_u}}=
\prod_{k=1}^{S}{\rm e}^{-i\gamma(t)\left(\omega_1\,s_2^{[k]}-\omega_2\,s_1^{[k]}\right)}\ ,
\label{4.33}
\end{equation}
where the function $\gamma(t)$ results from the time integration of $\omega_1(t)$ and $\omega_2(t)$
in (\ref{4.24}); explicitly:
\begin{equation}
\gamma(t)=\cosh(c\,\xi)\,\Big[\arctan\Big({\rm e}^{-\xi(t+c)}\Big)-\arctan\Big({\rm e}^{-\xi c}\Big)\Big]
\ .
\label{4.34}
\end{equation}
According to the result of {\sl Theorem \ref{theorem3}} above, the unitary transformation $U^{(S)}_t$
is responsible for the limiting time evolution of any local observables involving
the $S$ selected chain sites. In particular, in the case of single-site spin operators,
$\vec{s}=(s_1,s_2,s_3)$, one finds, dropping the now superfluous label $S$:
\begin{equation}
U_t^\dagger \, \vec{s}\ U_t = \mathcal{M}_t^{(\vec{\omega})}\cdot \vec{s} \ ,
\label{4.35}
\end{equation}
where the $3\times 3$ matrix $\mathcal{M}_t^{(\vec{\omega})}$ is explicitly given by:
\bigskip
\begin{equation}
\hskip -2.3cm
\mathcal{M}_t^{(\vec{\omega})}=\frac{1}{\omega_{12}}\pmatrix{
\omega_1^2\cos\gamma_{12}(t)+\omega_2^2&\omega_1\omega_2\left(\cos\gamma_{12}(t)-1\right)&\omega_{12}\,\omega_1\sin\gamma_{12}(t)\cr
\omega_1\omega_2\left(\cos\gamma_{12}(t)-1\right)&\omega_2^2\cos\gamma_{12}(t)+\omega_1^2&\omega_{12}\,\omega_2\sin\gamma_{12}(t)\cr
-\omega_{12}\,\omega_1\sin\gamma_{12}(t)&-\omega_{12}\,\omega_2\sin\gamma_{12}(t)&\omega_{12}^2\cos\gamma_{12}(t)
}\ ,
\label{4.36}
\end{equation}
\bigskip
with $\omega_{12}=\sqrt{(\omega_1)^2 + (\omega_2)^2}$ and $\gamma_{12}(t)=\omega_{12}\,\gamma(t)$.

\subsection{Dissipative dynamics of fluctuations: emergent non-linear open dynamics}
\label{section4.3}

After having studied the large $N$ dynamics dictated by the microscopic dissipative
evolution equation (\ref{4.4}) on quasi-local observables,
we shall now consider the limiting dynamics of fluctuation operators.

As explained in Section \ref{section2.4}, they form an algebra, the fluctuation algebra,
which is determined by
choosing the set of relevant single-particle, hermitian observables
generating the linear span $\mathcal{X}$ as in (\ref{2.16}). 
In the present case, it is natural to focus on the basis elements $v_\mu\in \mathfrak{a}$
entering the generator $\mathbb{L}^{(N)}$ through the Hamiltonian $H^{(N)}$ in (\ref{4.2}) 
and the dissipative contribution $\mathbb{D}^{(N)}$ in (\ref{4.6}), so that
\begin{equation}
\mathcal{X}=\Big\{v_r\ \big|\ v_r\equiv\vec{r}\cdot\vec{v}=\sum_{\mu=1}^{d^2} r_\mu\, v_\mu,\ 
\vec{r}\in\mathbb{R}^{d^2}\Big\}\ .
\label{4.37}
\end{equation}
The definition given in (\ref{2.14}) of the fluctuation operators needs however to be modified,
as the chosen system state $\omega$  need not be left
invariant by the microscopic time evolution generated by $\mathbb{L}^{(N)}$:
\begin{equation}
\omega_t^{(N)}=\omega\circ\Phi_t^{(N)}\neq\omega\ .
\label{4.38}
\end{equation}
As fluctuations account for deviations of observables from their mean values,
it is then necessary to extend the definition (\ref{2.14}) to a time-dependent one:
\begin{equation}
F^{(N)}_t(v_\mu)\equiv \frac{1}{\sqrt N}\sum_{k=1}^N \Big(v_\mu^{[k]}
-\omega_t^{(N)}\left(v_\mu^{[k]}\right){\bf 1}\Big)\ .
\label{4.39}
\end{equation}
This change guaranties the vanishing of the mean value of fluctuations,
$\omega_t^{(N)}\Big(F^{(N)}_t(v_\mu)\Big)=\,0$, a property that needs to be satisfied for all times.
The new definition (\ref{4.39}) further implies that also the symplectic matrix 
defined in (\ref{2.21}) might be in general time-dependent:
\begin{equation}
\sigma^{(\vec{\omega}_t)}_{\mu\nu}
=-i\lim_{N\to\infty} \omega_t^{(N)}\Big( \Big[F_t^{(N)}(v_\mu),\ F_t^{(N)}(v_\nu)\Big]\Big)\ .
\label{4.40}
\end{equation}
However, its dependence on time occurs only through the vector $\vec{\omega}_t$ of mean-field averages
introduced in (\ref{4.9}); indeed, the commutator:
\begin{equation}
\Big[F_t^{(N)}(v_\mu),\ F_t^{(N)}(v_\nu)\Big]=\frac{1}{N}\sum_{k=1}^N \left[v_\mu^{[k]},\ v_\nu^{[k]}\right]
=\frac{1}{N}\sum_{k=1}^N\sum_{\alpha=1}^{d^2} f_{\mu\nu}{}^\alpha v_\alpha^{[k]}\ ,
\label{4.41}
\end{equation}
results time-independent, while:
\begin{equation}
\sigma^{(\vec{\omega}_t)}_{\mu\nu}=-i\sum_{\alpha=1}^{d^2} f_{\mu\nu}{}^\alpha\, \omega_\alpha(t)\ .
\label{4.42}
\end{equation}

Let us now consider the fluctuation operators corresponding to the generic combination
$v_r\in\mathcal{X}$ at time $t=\,0$; dropping for simplicity the superfluous label $0$,
one then has ({\it cf.} (\ref{2.17})):
\begin{equation}
F^{(N)}(v_r)=\sum_{\mu=1}^N r_\mu\, F^{(N)}(v_\mu)\equiv \vec{r}\cdot\vec{F}^{(N)}(v)\ ,
\label{4.43}
\end{equation}
together with the corresponding Weyl-like operators, as in (\ref{2.29}):
\begin{equation}
W^{(N)}(\vec{r}\,)\equiv \e^{i\vec{r}\cdot\vec{F}^{(N)}(v)}\ .
\label{4.44}
\end{equation}
For a state $\omega$ which satisfy the two properties in (\ref{2.31}) and (\ref{2.32}),
in the large $N$ limit, $W^{(N)}(\vec{r}\,)$ give rise to Weyl operators:
\begin{equation}
\lim_{N\to\infty} W^{(N)}(\vec{r}\,) = W(\vec{r}\,)=\e^{i\vec{r}\cdot\vec{F}}
\ ;
\label{4.45}
\end{equation}
they are elements of the Weyl algebra $\mathcal{W}(\mathcal{X},\sigma^{(\vec{\omega})})$ defined with
the symplectic matrix $\sigma^{(\vec{\omega})}$, with components as in (\ref{4.42}), but evaluated at $t=\,0$.
Indeed, the bosonic operators $F_\mu$,
\begin{equation}
\lim_{N\to\infty}F^{(N)}(v_\mu)=F_\mu\ ,\qquad \mu=1,2,\ldots,d^2\ ,
\label{4.46}
\end{equation}
obey the following commutator relations: 
$\left[F_\mu,\,F_\nu\right]=i\sigma_{\mu\nu}^{(\vec{\omega})}$. As discussed in Section \ref{section2}, these limits
need to be understood as mesoscopic limits,
\begin{equation}
\lim_{N\to\infty}\omega\Big( W^{(N)}(\vec{r}\,)\Big)=
\e^{-\frac{1}{2} \vec{r}\cdot\Sigma^{(\vec{\omega})}\cdot \vec{r}}=
\Omega\Big( W(\vec{r}\,)\Big)\ ,\qquad  \vec{r}\in\mathbb{R}^{d^2}\ ,
\label{4.47}
\end{equation}
where $\Omega$ is the Gaussian state on the algebra $\mathcal{W}(\mathcal{X},\sigma^{(\vec{\omega})})$ defined
by covariance matrix:
\begin{equation}
\Sigma^{(\vec{\omega})}_{\mu\nu}=
\frac{1}{2}\lim_{N\to\infty} \omega\Big( \Big\{F^{(N)}(v_\mu),\ F^{(N)}(v_\nu)\Big\}\Big)\ .
\label{4.48}
\end{equation}

We are now ready to study the behaviour in the limit of large $N$ of the microscopic dissipative dynamics $\Phi_t^{(N)}$
generated by the dissipative generator $\mathbb{L}^{(N)}$ in (\ref{4.4}) on
the Weyl-like operators (\ref{4.44}). Recalling the definition (\ref{2.37}), one can show that:
\begin{equation}
\hskip -1.6 cm
\lim_{N\to\infty} \omega\Big( W^{(N)}(\vec{r}_1)\,\Phi^{(N)}_t\big[W_t^{(N)}(\vec{r}\,)\big]\, 
W^{(N)}(\vec{r}_2)\Big)=
\Omega\Big( W(\vec{r}_1)\,\Phi_t^{(\vec{\omega})}\big[W(\vec{r}\,)\big]\, W(\vec{r}_2)\Big)\ ,
\label{4.49}
\end{equation}
for all $\vec{r},\ \vec{r}_{1},\ \vec{r}_{2}\in \mathbb{R}^{d^2}$,
where $W_t^{(N)}(\vec{r}\,)$
is the Weyl-like operator constructed with the time-dependent fluctuation 
operator introduced in (\ref{4.39}), $W_t^{(N)}(\vec{r}\,)=\e^{i\vec{r}\cdot\vec{F}_t^{(N)}}$.
This limit defines the
mesoscopic dynamics $\Phi_t^{(\vec{\omega})}$ 
on the Weyl algebra $\mathcal{W}(\mathcal{X},\sigma^{(\vec{\omega})})$,
whose explicit form is given by the following result \cite{Carollo5}:

\begin{theorem}
\label{theorem 4}
The dynamics of quantum fluctuations is given by 
the mesoscopic map $\Phi_t^{(\vec{\omega})}\equiv m-\lim_{N\to\infty}\Phi_t^{(N)}$, where
\begin{equation}
\Phi_t^{(\vec{\omega})}\Big[W(\vec{r}\,)\Big]=\e^{-\frac{1}{2}\vec{r}\cdot\, Y_t^{(\vec{\omega})}\cdot\vec{r}}\
W(\vec{r}_t)\ ,\qquad \vec{r}_t=\left(X_t^{(\vec{\omega})}\right)^T\cdot \vec{r}\ ,
\label{4.50}
\end{equation}
with, 
\begin{eqnarray}
\label{4.51-1}
&&\hskip -1.5cm 
X_t^{(\vec{\omega})}=\mathbb{T}{\rm e}^{\int_0^t{\rm d}s\, Q^{(\vec{\omega}_s)}}\\
\label{4.51-2}
&&\hskip -1.5cm
Q^{(\vec{\omega}_t)}=-i\sigma^{(\vec{\omega}_t)}\,\widetilde{B}\,+\,D^{(\vec{\omega}_t)}\\
\label{4.51-3}
&&\hskip -1.5cm
Y_t^{(\vec{\omega})}=X_t^{(\vec{\omega})}\cdot\Bigg[\int_0^t{\rm d}s\, (X_s^{(\vec{\omega})})^{-1}\,
\Big(\sigma^{(\vec{\omega}_s)}\cdot A\cdot [\sigma^{(\vec{\omega}_s)}]^T\Big)\,[(X_s^{(\vec{\omega})})^{-1}]^T\Bigg]
\cdot (X_t^{(\vec{\omega})})^T\ .
\end{eqnarray}
\end{theorem}
In the above expression, $\widetilde{B}=B+\,2i\,h$ as defined in (\ref{4.12-3}), while
$A$ and $B$ are the symmetric and antisymmetric components of the Kossakowski matrix $C$ ({\it cf.} (\ref{4.11}));
further, $D^{(\vec{\omega}_t)}$ is the matrix defined in (\ref{4.14}), while $\sigma^{(\vec{\omega}_t)}$ 
is the time-dependent symplectic matrix with entries given by (\ref{4.42}).

The structure of the mesoscopic dynamics looks like that of Gaussian maps transforming Weyl operators 
into Weyl operators with rotated parameters and further multiplied by a damping factor; note in fact
that $Y_t^{(\vec{\omega})}$ is positive, since so is the Kossakoski matrix, hence $A$.
However, its explicit dependence on the 
mean-field quantities $\vec{\omega}$ makes the maps $\Phi^{(\vec{\omega})}_t$ 
not respectful of the algebraic structure of the Weyl algebra $\mathcal{W}(\mathcal{X},\sigma^{(\vec{\omega})})$,
making them acting non-linearly on it. 

Indeed, let us consider the action of $\Phi^{(\vec{\omega})}_t$ on the product of two Weyl operators.
Assuming linearity,  using the Weyl algebraic relations (\ref{2.22}), one would write:
$$
\Phi^{(\vec{\omega})}_t\left[W(\vec{r_1})W(\vec{r}_2)\right]=
\Phi^{(\vec{\omega})}_t\left[{\rm e}^{i\,\vec{r}_2\cdot\sigma^{(\vec{\omega})}\cdot\vec{r}_1}
\,W(\vec{r_2})W(\vec{r}_1)\right]={\rm e}^{i\,\vec{r}_2\cdot\sigma^{(\vec{\omega})}\cdot\vec{r}_1}
\,\Phi^{(\vec{\omega})}_t\left[\,W(\vec{r_2})W(\vec{r}_1)\right]\ .
$$ 
However, direct evaluation gives instead:
\begin{equation}
\Phi^{(\vec{\omega})}_t\left[W(\vec{r}_1)W(\vec{r}_2)\right]
={\rm e}^{i\,\vec{r}_2\cdot\sigma^{(\vec{\omega}_t)}\cdot\vec{r}_1}
\,\Phi^{(\vec{\omega})}_t\left[\,W(\vec{r}_2)W(\vec{r}_1)\right]\ ,
\label{4.52}
\end{equation}
where the symplectic matrix appearing in the prefactor is $\sigma^{(\vec{\omega}_t)}$
and not the one at $t=\,0$.
This is a consequence of the fact that the local operators $W^{(N)}(\vec{r_1})$ and $W^{(N)}(\vec{r_2})$ 
satisfy a Baker-Campbell-Haussdorf relation of the form:
$$
W^{(N)}(\vec{r}_1)\,W^{(N)}(\vec{r}_2)=W^{(N)}(\vec{r}_2)\,W^{(N)}(\vec{r}_1)\,
\exp\left(\Big[\vec{r}_2\cdot\vec{F}^{(N)}\,,\,\vec{r}_1\cdot\vec{F}^{(N)}\Big]\,
+\,O\left(\frac{1}{\sqrt{N}}\right)\right)\ .
$$
Since the leading order term in the argument of the exponential function is a mean-field quantity, 
it keeps evolving in time under the action of $\Phi^{(N)}_t$,
becoming the scalar quantity $i\,\vec{r}_2\cdot\sigma^{(\vec{\omega}_t)}\cdot\vec{r}_1$ in the large $N$ limit.

At first sight, the non-linearity of the obtained mesoscopic dynamics on the Weyl algebra 
$\mathcal{W}(\mathcal{X},\sigma^{(\vec{\omega})})$ appears rather puzzling, as any physically consistent
generalized quantum dynamics should be described by a semigroup of linear, completely positive maps.
The origin of this apparent clash stems from the explicit dependence on time of the
symplectic matrix, leading to time-evolving canonical commutation relations,
a rather uncommon situation. 
The proper tool to deal with such instances is provided by a suitable algebra extension,
allowing to deal with quantum fluctuations obeying algebraic rules that depend on the macroscopic averages.
One is thus led to introduce a hybrid system, in which there appear together quantum and classical degrees 
of freedom, strongly intertwined since the commutator of two fluctuations is a classical dynamical variable. 

Without entering into technical details (see \cite{Carollo5} for the full treatment), the dynamical
maps $\Phi^{(\vec{\omega})}_t$ can be extended to linear maps $\Phi_t$ on a larger algebra
than $\mathcal{W}(\mathcal{X},\sigma^{(\vec{\omega})})$. 
The Weyl algebra  $\mathcal{W}(\mathcal{X},\sigma^{(\vec{\omega})})$ explicitly depends
on the the vector $\vec{\omega}$ of macroscopic averages
through the symplectic matrix $\sigma^{(\vec{\omega})}$;
the idea is then to collect together these algebras for all possible values of
$\vec{\omega}$. The proper mathematical way to due this is through a direct integral
von Neumann algebra \cite{Bing}:
\begin{equation}
\mathcal{W}(\mathcal{X})\equiv\int^{\oplus}{\rm d}\vec{\omega}\ \mathcal{W}(\mathcal{X},\sigma^{(\vec{\omega})})\ .
\label{4.53}
\end{equation}
The most general element of this extended algebra $\mathcal{W}(\mathcal{X})$ 
are operator-valued functions $W^f_{\vec{r}}$, defined by:
\begin{equation}
W^f_{\vec{r}}:\, \vec{\omega}\mapsto f(\vec{\omega})\,W^{(\vec{\omega})}(\vec{r}\,)\ ,
\label{4.54}
\end{equation}
where $f$ is any element of the von Neumann algebra of bounded functions 
with respect to the measure ${\rm d}\vec{\omega}$, 
while $W^{(\vec{\omega})}(\vec{r}\,)$ is a Weyl operators in $\mathcal{W}(\mathcal{X},\sigma^{(\vec{\omega})})$,
{\it i.e.} the operator-valued functions $W^1_{\vec{r}}$ evaluated at 
$\vec{\omega}$.

On this extended algebra $\mathcal{W}(\mathcal{X})$, one can consider the action of a linear dynamical map $\Phi_t$
defined as follows:
\begin{equation}
\left(\Phi_t\Big[W^f_{\vec{r}}\Big]\right)(\vec{\omega})=
f(\vec{\omega}_t)\,\Phi^{(\vec{\omega})}_t\left[W^{(\vec{\omega})}(\vec{r}\,)\right]\ .
\label{4.55}
\end{equation}
One can show that these extended maps $\Phi_t$ form a one-parameter semigroup of completely positive, 
unital, Gaussian maps on the von Neumann algebra $\mathcal{W}(\mathcal{X})$.

The generator $\mathbb{L}$ of this semigroup can be obtained in the usual way
by taking the time-derivative of $\Phi_t$ at $t=0$; clearly, because of the direct integral form
of the algebra $\mathcal{W}(\mathcal{X})$ on which it acts, it will be of the form 
$\mathbb{L}=\int^\oplus{\rm d}\vec{\omega}\ \mathbb{L}^{(\vec{\omega})}$. 
The components $\mathbb{L}^{(\vec{\omega})}$ of the generator result of hybrid form 
\cite{Ciccotti}-\cite{Kapral},
containing a drift contribution that makes $\vec{\omega}$ evolve in time 
as a solution to the dynamical equation (\ref{4.14}),
together with mixed classical-quantum pieces and fully quantum contributions.
As such, it can not be written in the typical Kossakowski-Lindblad form; actually, in general, even the
purely quantum contributions can not be cast in this form, despite the fact that the linear maps
$\Phi_t$ constitute a semigroup of completely positive transformations.

These results have been only recently clarified and their applications to concrete
situations are in the process of being developed; they are not only of mathematical interest, 
but also of great physical relevance since in almost all experimental setups 
the macroscopic properties of the system actually vary in time.
This observation is particularly important in applications in quantum information 
and communication protocols based on collective bosonic degrees of freedom
requiring the presence of quantum correlations. Since non-classical correlations
({\it e.g.} entanglement), are directly related to the behaviour of the commutation relations, 
the hybrid dynamical structure presented above may play an important role
in modelling actual experiments.

As a preliminary step in this direction, in the next Section we shall see that, in analogy 
with the results presented in Section \ref{section3.3}, entanglement can be dissipatively generated at the mesoscopic level
of quantum fluctuations also by starting with microscopic dynamics generated by mean-field operators
of the form (\ref{4.6}).

\subsection{Mesoscopic entanglement through dissipation: mean-field dynamics}
\label{section4.4}

Let us reconsider the many-body system composed by two spin-1/2 chains and immersed in a common
environment introduced in Section \ref{section3.3.1}. As discussed there, the single-particle algebra is given by
$\mathfrak{a}=\mathcal{M}_2(\mathbb{C})\otimes \mathcal{M}_2(\mathbb{C})$ and the sixteen tensor
products $s_i\otimes s_j$, $i,j=0,1,2,3$, built with the spin operators 
$s_1$, $s_2$, $s_3$ and $s_0={\bf 1}/2$, constitute a basis in it. 
We shall equip the system with the state $\omega=\bigotimes_k\ \omega^{[k]}_\zeta$,
tensor product of the same single-site state $\omega_\zeta$ for all sites,
for which the only nonvanishing single-site expectations are:
\begin{equation}
\omega_\zeta\left(s_3\otimes {\bf 1}\right)=
\omega_\zeta\left({\bf 1}\otimes s_3\right)=-\zeta\ ,\quad
\omega_\zeta\left(s_3\otimes s_3\right)=-\zeta^2\ ,\quad \zeta\geq0\ .
\label{4.56}
\end{equation}

We are interested in the dissipative effects induced by the environment on the many-body system
at the mesoscopic level by a microscopic dynamics of mean field type; we shall thus neglect
any Hamiltonian contribution and
focus on a microscopic time evolution generated by an operator of the form (\ref{4.6}).
Further, instead of dealing with all the sixteen operators $s_i\otimes s_j$, it suffices
to restrict the treatment to a set $\mathcal{X}$ generated by the following six
single-site operators:
\begin{eqnarray}
\label{4.56-1}
&&v_1=s_1\otimes s_0\ ,\quad
v_2=s_2\otimes s_0\ ,\quad v_3=s_3\otimes s_0\ ,\\
\label{4.56-2}
&&v_4=s_0\otimes s_1\ , \quad v_5=s_0\otimes s_2\ ,\quad v_6=s_0\otimes s_3\ .
\end{eqnarray}
Notice that the three operators (\ref{4.56-1})
represent single-particle observables pertaining to the first chain, while the remaining three
refer to the second chain.

For a system composed by $N$ sites, out of these six single-site operators,
we can then construct the operators $V^{(N)}_\mu = \frac{1}{\sqrt N}\sum_{k=1}^N v^{[k]}_\mu$,
$\mu=1,2,\ldots,6$, scaling as fluctuations.  Given any quasi-local element $X$ of the system,
its microscopic dynamics will then be described by an evolution equation of the form:
$\partial_t X(t)=\mathbb{L}^{(N)}[X(t)]$; as generator, we take:
\begin{equation}
\mathbb{L}^{(N)}[X]=\sum_{\mu,\nu=1,2,4,5} C_{\mu\nu}\left(V^{(N)}_\mu\,X\,V^{(N)}_\nu\,
+\frac{1}{2}\,
\Big\{V^{(N)}_\mu\,V^{(N)}_\nu ,X\,\Big\}\right)\ ,
\label{4.57}
\end{equation}
involving only four operators $V^{(N)}_\mu$, and choose
a Kossakowski matrix $C$ of the form:
\begin{equation}
C=\pmatrix{ 1 & 1\cr
            1 & 1\cr} \otimes
   \pmatrix{ 1 & -ib\cr
   			ib & a\cr}\ ,\qquad  a\geq b^2\ .
\label{4.58}
\end{equation}

Let us first focus on the dynamics of the macroscopic observables, and as in Section \ref{section4.1},
study the large $N$ behaviour of the averages of the mean-field operators constructed with the
six basis elements of $\mathcal{X}$, 
$\omega_\mu(t):=\lim_{N\to\infty}\omega\left(\Phi_t^{(N)}\left[\frac{1}{N}\sum_{k=1}^{N}v_\mu^{[k]}\right]\right)$,
$\mu=1,2,\ldots,6$, with $\Phi_t^{(N)}=\e^{t \mathbb{L}^{(N)}}$.
Their evolution is given by the nonlinear equations in (\ref{4.13}); since in the present case the structure
constant $f$ are given by the $\epsilon$ symbol, see (\ref{2.38}), and $\tilde{B}$ reduces 
to the antisymmetric part of the Kossakowski matrix in (\ref{4.58}), one explicitly finds:
\begin{eqnarray}
\nonumber
&&\frac{{\rm d}}{{\rm d}t}\mathfrak{s}_1=-b\, \mathfrak{s}_1\, \mathfrak{s}_3 -b\, \mathfrak{s}_3\, \mathfrak{t}_1\ ,\\
\nonumber
&&\frac{{\rm d}}{{\rm d}t}\mathfrak{s}_2=-b\, \mathfrak{s}_2\, \mathfrak{s}_3 -b\, \mathfrak{s}_3\, \mathfrak{t}_2\ ,\\
\label{4.59}
&&\frac{{\rm d}}{{\rm d}t}\mathfrak{s}_3=b\, (\mathfrak{s}_1)^2 + b\, (\mathfrak{s}_2)^2 +
b\, \mathfrak{s}_1\, \mathfrak{t}_1 +b\, \mathfrak{s}_2\, \mathfrak{t}_2\ ,\\
\nonumber
&&\frac{{\rm d}}{{\rm d}t}\mathfrak{t}_1=-b\, \mathfrak{t}_1\, \mathfrak{t}_3 -b\, \mathfrak{t}_3\, \mathfrak{s}_1\ ,\\
\nonumber
&&\frac{{\rm d}}{{\rm d}t}\mathfrak{t}_2=-b\, \mathfrak{t}_2\, \mathfrak{t}_3 -b\, \mathfrak{t}_3\, \mathfrak{s}_2\ ,\\
\nonumber
&&\frac{{\rm d}}{{\rm d}t}\mathfrak{t}_3=b\, (\mathfrak{t}_1)^2 + b\, (\mathfrak{t}_2)^2 +
b\, \mathfrak{t}_1\, \mathfrak{s}_1 +b\, \mathfrak{t}_2\, \mathfrak{s}_2\ ,
\end{eqnarray}
where, for sake of clarity, the components $\omega_\mu(t)$, $\mu=1,2,\ldots,6$, of the vector $\vec{\omega}(t)$
have been relabeled as
$\vec{\omega}=(\mathfrak{s}_1,\mathfrak{s}_2,\mathfrak{s}_3, \mathfrak{t}_1,\mathfrak{t}_2,\mathfrak{t}_3)$.

Recalling that the chosen state $\omega$ for the system satisfies the properties (\ref{4.56}), one immediately sees
that the initial conditions for this nonlinear system of equations at $t=\,0$ are:
$\vec{\omega}=(0,0,-\zeta,0,0,-\zeta)$. But this is a fixed point of the system (\ref{4.59}), so that in this
particular case, the macroscopic observables are time-independent, or equivalently,
the microscopic state $\omega$ is left invariant by the evolution generated by (\ref{4.57}).

Using these results, one can now study the limiting dynamics of the fluctuation operators
constructed out of the single-site observables (\ref{4.56-1}) and (\ref{4.56-2}),
or equivalently of the elements of their linear span $\mathcal{X}$. 
Since the macroscopic observables result time-independent, the generalized definition of fluctuations
in (\ref{4.39}) reduces to the original one in (\ref{2.14}), without any time dependence in the
averaged term. The fluctuation operators are then defined by:
\begin{equation}
F^{(N)}(v_\mu)\equiv \frac{1}{\sqrt{\zeta N}}\sum_{k=1}^N \Big(v_\mu^{[k]}-\omega\left(v_\mu^{[k]}\right)\Big)\ ,
\qquad \mu=1,2,\ldots, 6\ ,
\label{4.60}
\end{equation}
where, for later convenience, we have also included a rescaling factor $1/\sqrt\zeta$,
while their corresponding Weyl-like operators are given by:
\begin{equation}
W^{(N)}(\vec{r}\,)\equiv \e^{i\vec{r}\cdot\vec{F}^{(N)}(v)}\ , \qquad
\vec{r}\cdot\vec{F}^{(N)}(v)=\sum_{\mu=1}^6 r_\mu\, F^{(N)}(v_\mu)\ ,\quad \vec{r}\in\mathbb{R}^6\ .
\label{4.61}
\end{equation}

Since the chosen system state $\omega$ is translation invariant and manifestly satisfies
the clustering condition (\ref{2.12}), in the large $N$ limit
the Weyl-like operators (\ref{4.61}) define elements $W(\vec{r}\,)=\e^{i\vec{r}\cdot\vec{F}}$ of the Weyl algebra
$\mathcal{W}(\mathcal{X},\, \sigma^{(\omega)})$,
\begin{equation}
\lim_{N\to\infty}\omega\Big( W^{(N)}(\vec{r}\,)\Big)=
\e^{-\frac{1}{2} \vec{r}\cdot\Sigma^{(\omega)}\cdot \vec{r}}=
\Omega\Big( W(\vec{r}\,)\Big)\ ,\qquad  \vec{r}\in\mathbb{R}^6\ ,
\label{4.62}
\end{equation}
where $\sigma^{(\omega)}$ is the symplectic matrix as defined in (\ref{2.20}), explicitly giving
\begin{equation}
\sigma^{(\omega)}=\pmatrix{1 & 0\cr
						   0 & 1\cr}
\otimes
\pmatrix{ 0 & 1 & 0\cr
	     -1 & 0 & 0\cr
	      0 & 0 & 0\cr}\ ,
\label{4.63}
\end{equation}
while $\Omega$ is the Gaussian state on the algebra $\mathcal{W}(\mathcal{X},\, \sigma^{(\omega)})$
with covariance matrix $\Sigma^{(\omega)}$ as given in (\ref{2.21}),
\begin{equation}
\Sigma^{(\omega)}=\frac{1}{4\zeta}\, {\bf 1}_6\ .
\label{4.64}
\end{equation}
The Bose fields $F_\mu$ appearing in the Weyl operators $W(\vec{r}\,)$
are the mesoscopic limit of the fluctuation operators:
$\lim_{N\to\infty}F^{(N)}(v_\mu)=F_\mu$, for which:
$[F_\mu,\, F_\nu]=i\sigma^{(\omega)}_{\mu\nu}$. As a result, $F_3$ and $F_6$ are classical
variables, commuting with all remaining operators. 

We shall then focus on the reduced 
Weyl algebra $\mathcal{W}(\tilde\sigma^{(\omega)})$, with elements $W(\vec{r}\,)$
containing only the four operators $F_\mu$, $\mu=1,2,4,5$, and defined by the $4\times 4$
symplectic matrix $\tilde\sigma^{(\omega)}=i{\bf 1}_2\otimes\sigma_2$, obtained from
(\ref{4.63}) by deleting the third and sixth row/column. Similarly, the restriction 
$\widetilde\Omega$ of the state $\Omega$ on $\mathcal{W}(\tilde\sigma^{(\omega)})$ is
the two-mode Gaussian state with covariance $\widetilde\Sigma^{(\omega)}={\bf 1}_4/4\zeta$.
In fact, $(F_1,\, F_2)$ and $(F_4,\, F_5)$ constitute independent bosonic modes,
obeying standard commutation relation. In addition, because of the definitions (\ref{4.56-1})
and (\ref{4.56-2}), the first couple represents mesoscopic observables referring to the
first chain, while the second couple represents observables of the second chain.
It is then interesting to see whether these collective bosonic modes, pertaining to 
different chains, can get entangled
by the action of the limiting, mesoscopic dynamics obtained from the generator (\ref{4.57}).

As initial state we shall take the Gaussian state $\widetilde\Omega$, with covariance
$\widetilde\Sigma^{(\omega)}$ given above; being proportional to the unit matrix,
the state does not support any correlation (classical or quantal) between the two mesoscopic modes.
This state is not left invariant by the mesoscopic dynamics
$\Phi_t^{(\vec{\omega})}= m-\lim_{N\to\infty}\e^{t\mathbb{L}^{(N)}}$, explicitly given in
Theorem 4; indeed, one finds
that $\widetilde\Omega_t\equiv\widetilde\Omega\circ \Phi_t^{(\vec{\omega})}$ 
remains Gaussian, with a time-dependent covariance matrix given by:
\begin{equation}
\widetilde\Sigma^{(\omega)}_t= 
X_t^{(\vec{\omega})}\cdot \widetilde\Sigma^{(\vec{\omega})}\cdot \Big[X_t^{(\vec{\omega})}\Big]^T
+ Y_t^{(\vec{\omega})}\ ,
\label{4.65}
\end{equation}
with $X_t^{(\vec{\omega})}$ and $Y_t^{(\vec{\omega})}$ as in (\ref{4.51-1})-(\ref{4.51-3}).
Fortunately, in the present case the averages of macroscopic observables are time-independent, so that
these two matrices can be easily computed; explicitly:
\begin{equation}
X_t^{(\vec{\omega})}=\frac{1}{2} \pmatrix{ x^{(+)}_t & -x^{(-)}_t\cr
										  -x^{(-)}_t & x^{(+)}_t\cr} \otimes {\bf 1}_2\ ,
\qquad  x^{(\pm)}_t=1\pm \e^{-2b\zeta t}\ ,
\label{4.66}
\end{equation}
\begin{equation}
Y_t^{(\vec{\omega})}=y_t\, \pmatrix{ 1 & 1\cr
									 1 & 1\cr} \otimes
							\pmatrix{a & 0\cr
									 0 & 1\cr}\ ,
\qquad y_t=\frac{1}{4b}\left( 1- \e^{-4b\zeta t}\right)
\ .
\label{4.67}
\end{equation}

As discussed in Section \ref{section3.3.1}, the entanglement content of the evolved Gaussian state
can be studied by looking at the logarithmic negativity $E(t)$, defined in (\ref{3.37})
in terms of the smallest symplectic eigenvalue of the partially transposed
covariance matrix (\ref{4.65}).
One can check that there are regions in the $(a, b, \zeta)$ parameter space for which
$E(t)$ is indeed positive; further, one finds that the generated entanglement can persist for asymptotic
long times.

In order to show this, using (\ref{4.65})-(\ref{4.67}), let us compute the asymptotic covariance matrix 
$\widetilde\Sigma^{(\omega)}_\infty = \lim_{t\to\infty}\widetilde\Sigma^{(\omega)}_t$;
explicitly, one finds:
\begin{equation}
\widetilde\Sigma^{(\omega)}_\infty=\frac{1}{8\zeta} \pmatrix{ \Sigma^{(+)} & \Sigma^{(-)}\cr
															\Sigma^{(-)} & \Sigma^{(+)}\cr} \ ,
\label{4.68}
\end{equation}
with
\begin{equation}
\Sigma^{(\pm)}=\pmatrix{ 1 \pm \frac{2a\zeta}{b}  &  0\cr
						0				& 1 \pm \frac{2\zeta}{b}\cr}\ .
\label{4.69}
\end{equation}
With the help of these two matrices, one can now compute the asymptotic 
logarithmic negativity $E_\infty$ (see \cite{Souza,Isar}), obtaining:
\begin{equation}
E_\infty=-\log_2\left(\frac{1+a-|a-1|}{4b\zeta}\right) \ .
\label{4.70}
\end{equation}
For $a< 1$, and provided $a/2b\zeta<1$, $E_\infty$ is positive, thus signaling asymptotic entanglement between 
the two chains at the collective level of mesoscopic observables. Therefore, the microscopic
dissipative generator in mean-field form (\ref{4.57}) gives rise to a dynamical evolution at the level
of mesoscopic observables able to create quantum correlations among collective 
operators pertaining to different chains, and in addition to sustain this generated
entanglement for asymptotic long times.

\section{Outlook}
\label{section5}

The study of quantum many-body systems, {\it i.e.} of systems with a very large number $N$ of
microscopic constituents, requires analyzing collective observables, involving all system degrees of freedom.
Not all such collective operators are useful for discussing the quantum behaviour of the model,
since most of them lose any quantum character as the number of particles increases.
Mean-field observables are typical examples of this behaviour, as they form an abelian,
commutative algebra in the thermodynamic limit.

Only fluctuation-like operators, built out of deviations from mean values, do retain a quantum character
even in the large $N$ limit: these are the observables to be used for studying the behaviour
of many-body system at the mesoscopic level, in between the microscopic realm of their constituents
and the macroscopic, semiclassical scale. Quantum fluctuations turn out to be bosonic operators,
obeying canonical commutation relations. As the many-body system
is in general immersed in a weakly-coupled external environment, their dynamics is non-unitary,
encoding dissipative and noisy effects. It can be described by a one-parameter semigroup 
of completely positive maps generated by a master equation in Lindblad form,
although for interactions scaling as $1/N$, the so-called mean-field couplings, the semigroup
character of the time evolution can be recovered only through a suitable extension of the underlying
fluctuation algebra.

The presence of an external environment and the consequent dissipative phenomena it generates
usually lead to loss of quantum coherence. However, in certain circumstances, via a purely mixing mechanism,
the environment can act as a coherent enhancing medium for a couple of independent many-body systems immersed in it.
In these cases, mesoscopic entanglement between the two systems
can be generated at the level of quantum fluctuations. This result has clearly importance in
actual experiments, where {\it ab initio} preparation of many-body systems in an highly entangled state
is in general difficult; instead, inserting them in a suitably engineered environment could
more easily generate quantum correlations among them.

Finally, let us mention two additional developments of the theory of quantum fluctuation, not
included in the previous discussion. In systems with long-range correlations,
phase transitions could occur, so that the scaling of the order $1/\sqrt{N}$ might not be
appropriate in order to get physically sensible mesoscopic observables. In such cases,
one defines the so-called {\it abnormal} fluctuation operators $F^{(N)}_\delta$, scaling as $1/N^\delta$,
with $0<\delta<1$ \cite{Verbeure-book}. For states carrying a non-trivial second and third
moment for the observables $F^{(N)}_\delta$, one can show that the mesoscopic limit
$\lim_{N\to\infty} F^{(N)}_\delta=F_\delta$ defines well-behaved bosonic observables, belonging
to a non-abelian Lie algebra.

On the other hand, a different kind of canonical algebraic structure 
obeyed by quantum fluctuations has been discussed in Section \ref{section4} while
studying systems with mean-field like interactions, {\it i.e.} scaling as $1/N$. In general,
for such systems, the macroscopic observables explicitly depend on time, leading
to a modification of the definition of the fluctuation operators (see (\ref{4.39})).
At the mesoscopic level, this gives rise to limiting bosonic observables obeying commutation
relations evolving in time, providing an interesting instance in which algebraic and dynamical
features come out interconnected. Indeed, these mesoscopic, collective fluctuations
posses richer dynamical properties, able to reveal 
weak, but far reaching correlations between the system microscopic constituents \cite{Carollo4}; 
these correlations can not be detected by any local measure on the many-body system, 
but still have non-negligible effects on the dynamics of collective, fluctuation observables.

The presented results are just a selection of possible applications of quantum fluctuations
in modelling the collective quantum behaviour of open many-body system at the interface between the microscopic
and the macroscopic world; we are confident 
that our presentation will stimulate further theoretical developments,
as well as experimental applications.

%%
%\begin{equation}
%\ .
%\label{}
%\end{equation}
%%
%
%
%%
%\begin{eqnarray}
%\nonumber
%&&\ ,\\
%&&\ .
%\label{}
%\end{eqnarray}
%%

%\pagebreak

%\vfill\eject

\end{document}